\documentclass[amsmath,amssymb,aps,reprint]{revtex4-1}

\usepackage{graphicx}
\usepackage{dcolumn}
\usepackage{bm}
\usepackage{amssymb}
\usepackage{amsmath}
\usepackage{subfigure}
\usepackage{physics}
\usepackage{sublabel}
\usepackage[utf8]{inputenc}

\begin{document}

\title{Giant enhancement of Piezo-resistance in ballistic graphene due to transverse electric fields }

\author{Abhinaba Sinha}
\email[E-mail:~]{asinha@iitb.ac.in}
\affiliation{Department of Electrical Engineering, IIT Bombay, Powai, Mumbai-400076, India}
	
\author{Abhishek Sharma}
\email[E-mail:~]{sabhishek@iitb.ac.in}
\affiliation{Department of Electrical Engineering, IIT Bombay, Powai, Mumbai-400076, India}
	
\author{Ashwin Tulapurkar}
\affiliation{Department of Electrical Engineering, IIT Bombay, Powai, Mumbai-400076, India}
	
\author{V Ramgopal Rao}
\affiliation{Department of Electrical Engineering, IIT Bombay, Powai, Mumbai-400076, India}

\author{Bhaskaran Muralidharan}
\affiliation{Department of Electrical Engineering, IIT Bombay, Powai, Mumbai-400076, India}
	
\begin{abstract}

We investigate the longitudinal and transverse piezoresistance effect in suspended graphene in the ballistic regime. Utilizing parametrized tight binding Hamiltonian from ab initio calculations along with Landauer quantum transport formalism, we devise a methodology to evaluate the piezoresistance effect in 2D materials especially in graphene. We evaluate the longitudinal and transverse gauge factor of graphene along armchair and zigzag directions in the linear elastic limit ($0\%$-$10\%$). The longitudinal and transverse gauge factors are identical along armchair and zigzag directions. Our model predicts a significant variation ($\approx 1000\% $ change) in transverse gauge factor compared to longitudinal gauge factor along with sign inversion. The calculated value of longitudinal gauge factor is $\approx 0.3$ whereas the transverse gauge factor is $\approx -3.3$.
We rationalize our prediction using deformation of Dirac cone and change in separation between transverse modes due to longitudinal and transverse strain, leading to an inverse change in gauge factor. The results obtained herein may serve as a template for high strain piezoresistance effect of graphene in nano electromechanical systems. 

\end{abstract}
	
\maketitle

\section{Introduction}

Graphene became one of the most extensively researched material soon after its discovery in 2004. It is the first single atomic thick 2D material, isolated in the laboratory. Owing to its unique properties, often the terms wonder material \cite{Geim2009} and miracle material \cite{Novoselov2012} are assigned to it. These unique properties mostly originate from its hexagonal layered 2D structure. Graphene is one of the strongest known material due to the presence of strong planar bonds~\cite{Lee2008}. Thus, graphene can undergo elastic deformation for more than $ 20\% $ strain \cite{Kim2009,Liu2007}. Additionally, the out of plane $\pi$ electrons lead to very high electrical \cite{Novoselov2004} and thermal conductivity \cite{Baladin2008}. Due to the symmetry between the two inter-penetrating triangular sub-lattices, graphene has a zero band gap \cite{Novoselov2011,Deshmukh2011}. It also exhibits a linear dispersion relation at small energy. Consequently, electrons in graphene behave like relativistic particles \cite{CastroNeto2009,Novoselov2011}. At sub-micron length, graphene behaves like a ballistic conductor \cite{Novoselov2004,Das2008}. The combination of all these properties in a single material brings about various novel applications in the field of flexible electronics \cite{El-Kady2012,Georgiou2012,Das2018}, photodetectors \cite{xia2009,mueller2010,zhang2013,liu2014}, solar cells \cite{Li2010,Miao2012}, photonic devices \cite{bonaccorso2010,bao2012}, just to name a few. Besides these applications, graphene has applications in MEMS systems as sensors \cite{Smith2013,Dolleman2015,Wang2016}, switches \cite{Li2018}, resonators \cite{Chen2009}, actuators \cite{Huang2012,Rogers2011} etc.\\
\indent The high elastic limit of graphene is preferable for strain engineering applications. It enhances the range of operation for strain sensors. These sensors when combined with electrical or optical readouts, enable us to measure different physical quantities. A strain sensor that measures the change in resistance is known as a piezoresistance sensor. In this work, we restrict our discussion only to the piezoresistance effect in ballistic graphene. \\
\indent Graphene sheet exhibits 2D characteristics for width beyond a hundred nanometer \cite{Han2007}. Thus, graphene behaves as a 2D ballistic conductor for width more than $100~nm$ and less than its mean free path. \\
\indent Piezoresistance effect is measured by gauge factor (GF). GF is the normalized change in resistance with strain. The GF of graphene strongly depends on the type of graphene\cite{Huang2011,Zhu2013}, the substrate underneath\cite{Lee2010,Hosseinzadegan2012,Chen2011} and scattering mechanism involved\cite{Smith2013}. The GF of graphene in the ballistic regime is still not explored. Hence, in this work, we explore the longitudinal and transverse piezoresistance effect in graphene along armchair and zigzag directions in ballistic regime using quantum transport formalism.\\
\indent We develop a generic theoretical model for calculating the GF of 2D materials along different directions in ballistic regime and employed it on graphene. Our model computes GF from mode density using band counting method \cite{Jeong2009} and Landauer formalism along different directions.\\
\indent In subsequent sections, we describe the development of our mathematical model, calculate the transport properties and GF of graphene, and explain the underlying physics of the predicted value of longitudinal gauge factor (LGF) and transverse gauge factor (TGF) along armchair and zigzag directions. The detail derivation of mathematical expressions are discussed in Appendix.

\section{Theoretical Model } \label{section2}
\subsection {Simulation Setup}
\begin{figure}

	\subfigure[]{\includegraphics[height=0.17\textwidth,width=0.228\textwidth]{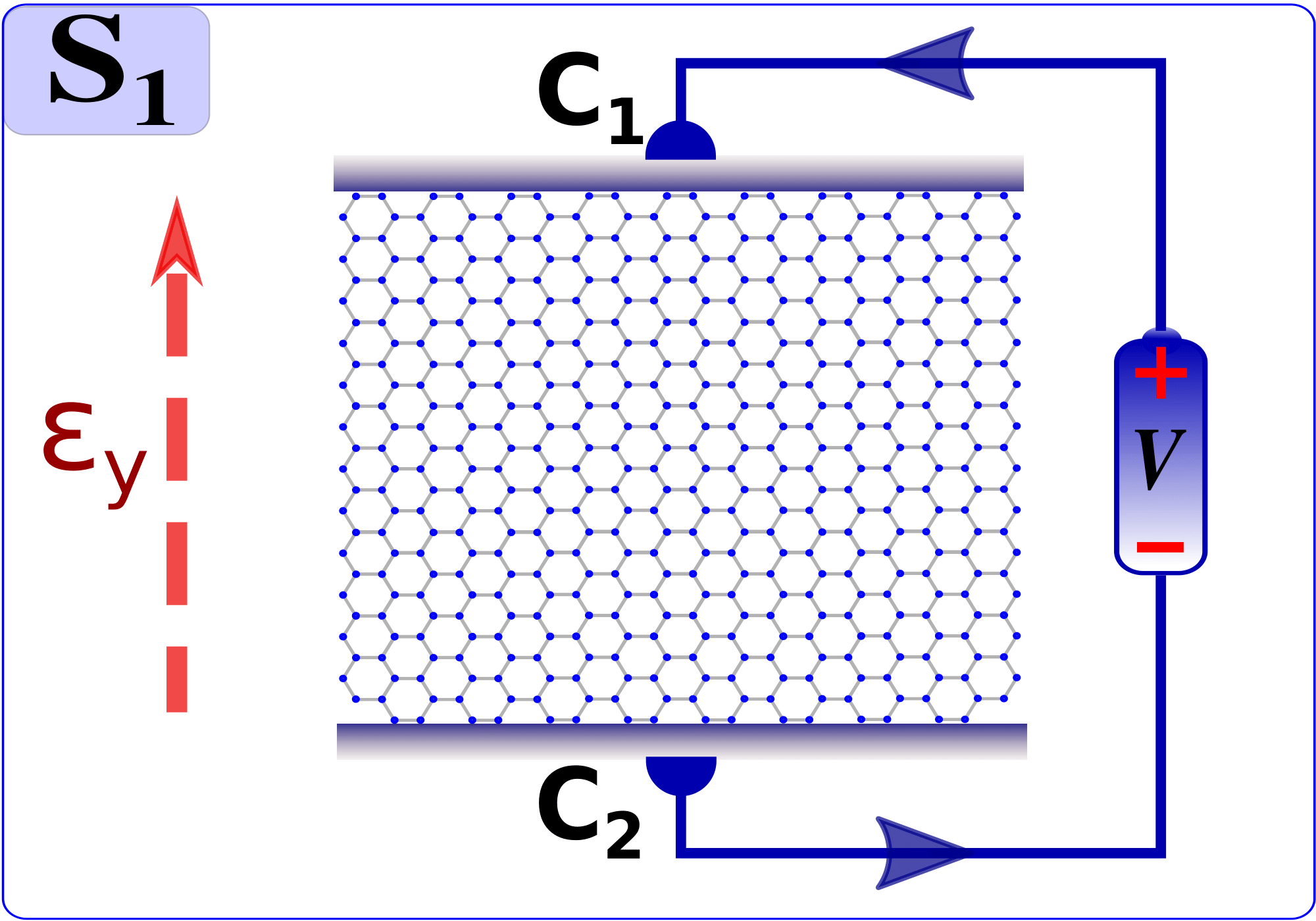}\label{s1}}
	\quad
	\subfigure[]{\includegraphics[height=0.17\textwidth,width=0.228\textwidth]{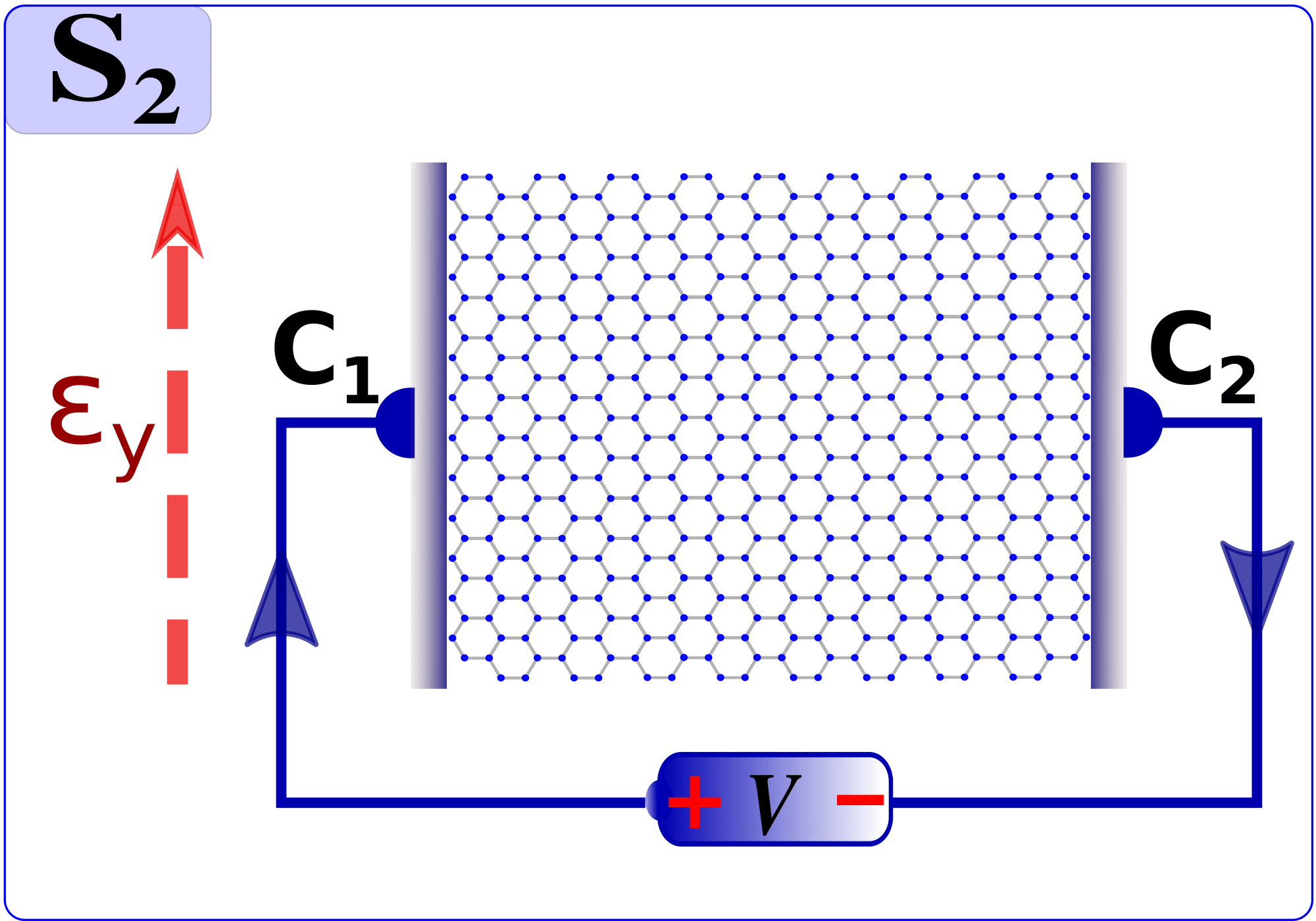}\label{s2}}
	\quad
	\subfigure[]{\includegraphics[height=0.20\textwidth,width=0.46\textwidth]{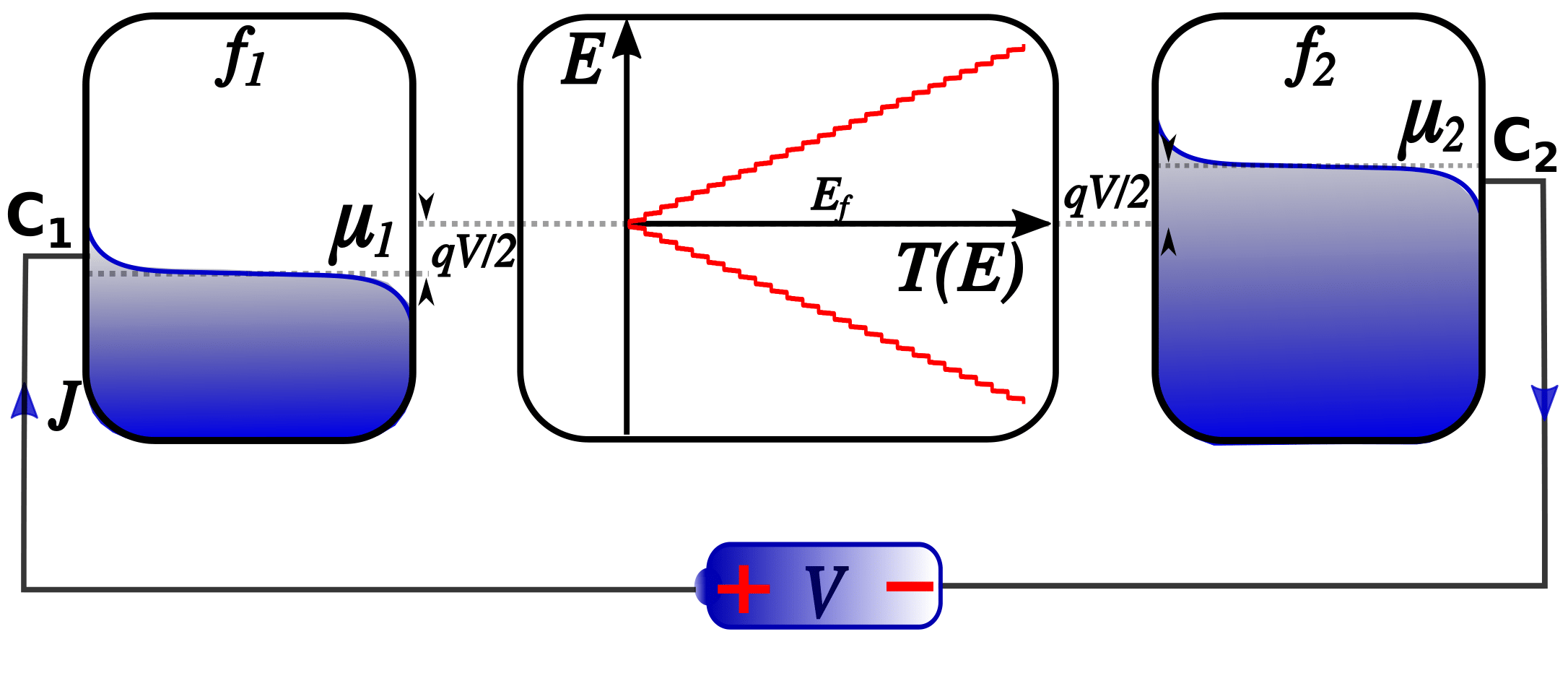}\label{transport-model}}
	\quad
	\caption{Schematic diagram of simulation setups $S_{1}$ and $S_{2}$ to determine the longitudinal and transverse GF of graphene. A uniaxial strain $\varepsilon_{y}$ is applied along the zigzag direction (y-axis) in both setups. (a) Setup $S_{1}$ consists of a voltage source (V) connected across the zigzag direction, using contacts $C_{1}$ and $C_{2}$.  Similarly, (b) setup $S_{2}$ consists of a voltage source (V) connected across the armchair direction, using contacts $C_{1}$ and $C_{2}$. (c) A generic quantum transport model for $S_{1}$ and $S_{2}$ is shown with contacts $C_{1}$ and $C_{2}$ connected across the graphene sheet having transmission T(E).}
	\label{Structure}
\end{figure}
\begin{figure}
  {\includegraphics[height=1\textwidth,width=0.45\textwidth]{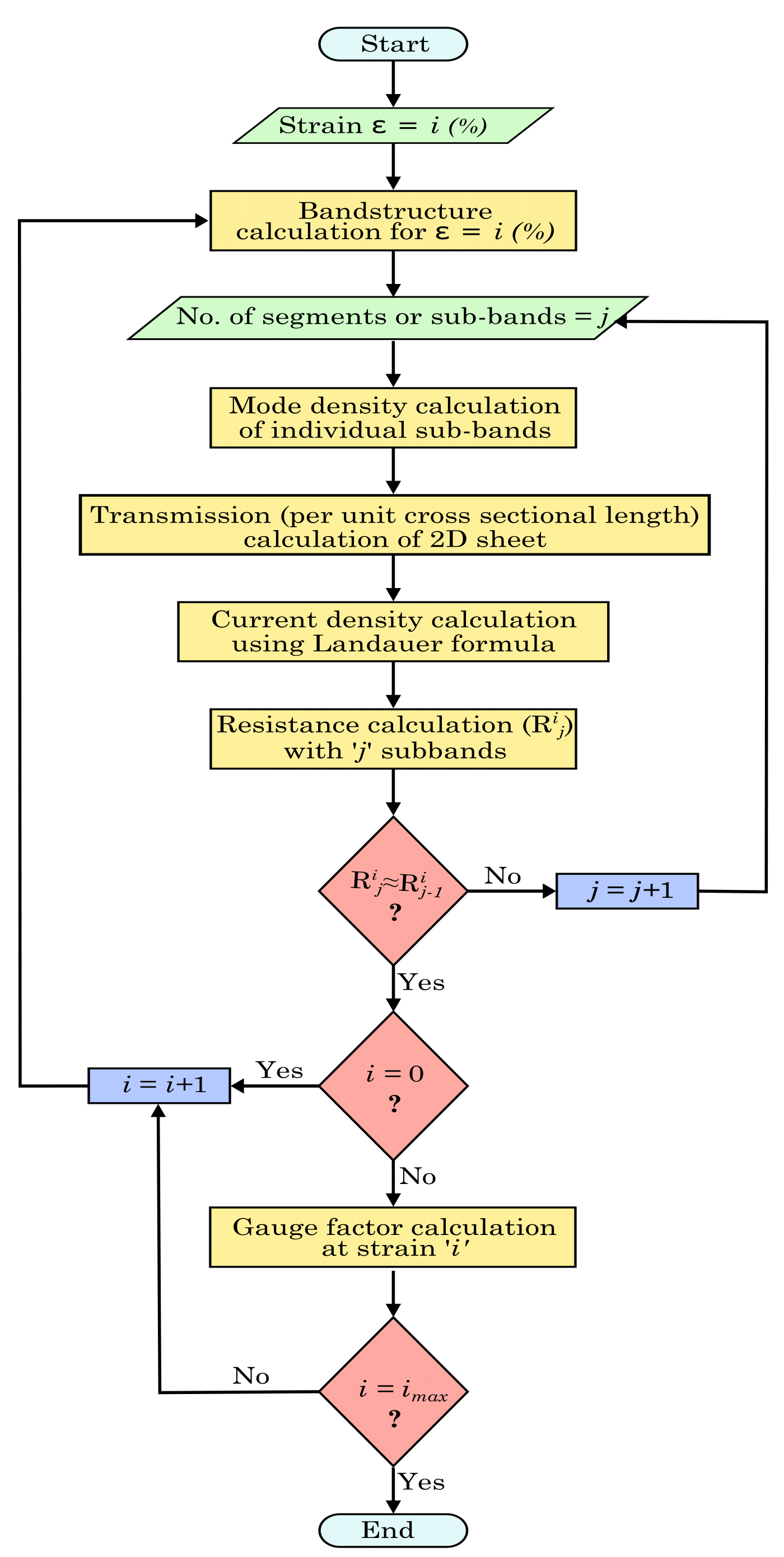}}
	\caption{Flow chart for gauge factor (GF) calculation of a 2D material in the ballistic regime. `$j$' is the number of segments used in band counting method, $R^{i}_{j}$ is resistance with `$j$' segments and $i_{max}$ is the maximum linear elastic limit of the 2D material. }  
	\label{Flowchart}
\end{figure}

\indent The setups $S_1$ and $S_2$ consist of a graphene sheet having contacts $C_{1}$ and $C_{2}$ across the zigzag direction (y-axis) and armchair direction (x-axis) as shown in Fig.~\ref{s1} and Fig.~\ref{s2} respectively. A uniaxial strain ($\varepsilon_{y}$) in the linear elastic regime ($0\%-10\%$)~\cite{Pereira2009} is applied along zigzag direction in $S_{1}$ and $S_{2}$. The magnitude of strain is gradually increased from $0\%$ to $10\%$ and simultaneously, the current density (J) is obtained for applied voltage (V) in the linear regime. The longitudinal gauge factor (LGF) is obtained from setup $S_{1}$ whereas the transverse gauge factor (TGF) is obtained from setup $S_{2}$ for zigzag direction. The LGF and TGF for armchair direction are also evaluated in a similar manner. \\
\indent The quantum transport model for $S_{1}$ and $S_{2}$ is shown in Fig.~\ref{transport-model}. The Fermi energy of the graphene channel ($E_{f}$) is at 0 eV. The Fermi function at $C_{1}$ is $f_{1}$ with fermi energy at $\mu_{1}=-qV/2$. Similarly, the Fermi function at $C_{2}$ is $f_{2}$ with Fermi energy at $\mu_{2}=qV/2$. For ease of calculation, armchair direction is taken along the x-axis and zigzag direction is taken along the y-axis.\\
\indent We sketch in Fig.~\ref{Flowchart}, a generic computational model that evaluates GF of 2D materials in the ballistic regime. We employ this model to compute the longitudinal and transverse GF of graphene along zigzag and armchair directions. Our model involves obtaining the Brillouin zone, getting the band structure of strained graphene using parametrized tight binding Hamiltonian, evaluation of the mode density function of graphene using the band counting method ~\cite{Jeong2009} and finally evaluation of GF using Landauer formalism. The detailed description of these steps are as follows:
\subsubsection{Brillouin zone and E-k relation of strained graphene}
\begin{figure}
	\subfigure[]{\includegraphics[height=0.20\textwidth,width=0.228\textwidth]{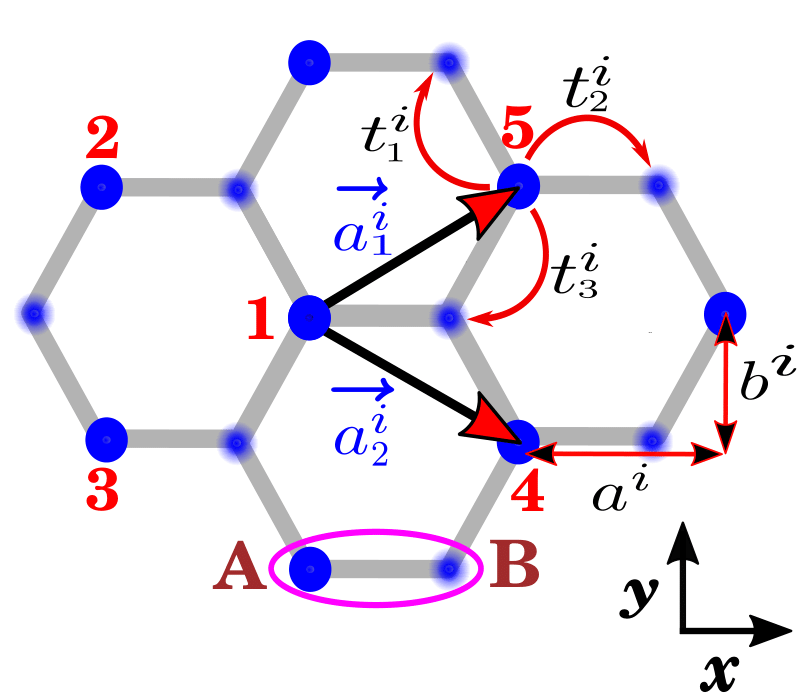}\label{rl}}
	\quad
	\subfigure[]{\includegraphics[height=0.20\textwidth,width=0.228\textwidth]{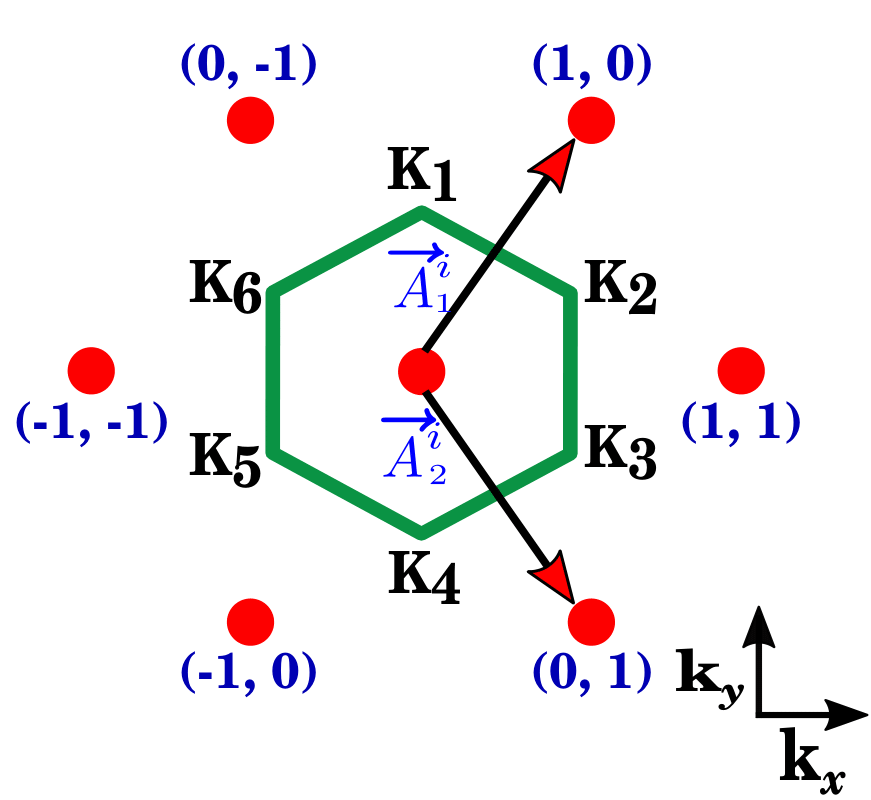}\label{bz}}
	\caption{(a) Real crystal lattice of strained graphene with $\vec{a^i_{1}} $, $\vec{a^i_{2}}$ as lattice vectors, A-B as basis and $t^{i}_1$, $t^{i}_2$ and $t^{i}_3$ as nearest neighbour tight binding parameters. (b) Reciprocal crystal lattice and $1^{st}$ Brillouin zone of strained graphene with reciprocal lattice vectors $ \vec{A^{i}_{1}}$ and $\vec{A^{i}_{2}}$. The intersection of the edges of Brillouin zone are labeled as $K_{1}$, $K_{2}$, ...., $K_{6}$ in clockwise manner and their corresponding Dirac points are denoted by $DP_{1}$, $DP_{2}$, ...., $DP_{6}$ respectively.} 
	\label{Lattice}
\end{figure}

Lattice vectors $\vec{a^i_{1}} $ and $\vec{a^i_{2}}$ describe the crystal-lattice of uniaxially strained graphene along armchair and zigzag directions. Superscript `$i$' denotes the magnitude of strain $\varepsilon_{x}$ and $\varepsilon_{y}$ in percentage. Figure \ref{rl} shows  schematic diagram of $\vec{a^i_{1}} $ and $\vec{a^i_{2}}$ in uniaxially strained graphene crystal. The lattice vectors $\vec{a^i_{1}} $ and $\vec{a^i_{2}}$ are given by:
\begin{equation}
\vec{a^i_{1}} = a^{i}\hat{x} + b^{i}\hat{y}
\label{eqa1}
\end{equation}
\begin{equation}
\vec{a^i_{2}} = a^{i}\hat{x} - b^{i}\hat{y}
\label{eqa2}
\end{equation}
where $ a^{i}=1.5 a_{0} (1+\epsilon_{x})$ and $b^{i}=(\sqrt3/2) a_{0}(1+\nu\epsilon_{x})$ in setup $S_{1}$, similarly $ a^{i}=1.5 a_{0} (1+\nu\epsilon_{y})$ and $b^{i}=(\sqrt3/2) a_{0}(1+\epsilon_{y})$ in setup $S_{2}$. Figure \ref{bz} shows the corresponding Brillouin zone of strained graphene. The reciprocal lattice vectors corresponding to Eq.~\eqref{eqa1} and Eq.~\eqref{eqa2} are given by:
\begin{equation}
\vec{A^{i}_{1}} =\frac{ 2\pi (\vec a^{i}_{2} \times \vec a^{i}_{3})}{ \vec a^{i}_1\cdot(\vec a^{i}_2 \times \vec a^{i}_3)}
\label{eqA1}
\end{equation}
\begin{equation}
\vec{A^{i}_{2}} = \frac{ 2\pi (\vec a^{i}_3 \times \vec a^{i}_1)}{ \vec a^{i}_2\cdot(\vec a^{i}_3 \times \vec a^{i}_1)}
\label{eqA2}
\end{equation}
 Graphene behaves elastically upto $20\%$ strain \cite{Liu2007,Kim2009,Lee2008}. We apply strain in the range of $0\%$ - $10\%$ that corresponds to the linear elastic regime in graphene~\cite{Pereira2009}. Poisson ratio ($\nu$) for uniaxial strain in graphene has been reported in the range of 0.10 - 0.20 \cite{Ribeiro2009,Liu2007,Farjam2009}. We use $\nu = -0.14$ and $a_{0}=1.42 \AA$ in all our calculations \cite{CastroNeto2009}.
$\vec{a^i_{3}}$ is taken as 1 for the ease of calculation. 
The corresponding reciprocal lattice points are given by:
\begin{equation}
\vec{G^{i}} = M \vec{A^i_{1}} + N \vec{A^i_{2}}
\label{eqG}
\end{equation}

The reciprocal lattice points nearest to the origin are shown with red dots in Fig.~\ref{bz}. A generic $1^{st}$ Brillouin zone for uniaxially strained graphene along armchair or zigzag direction is shown in Fig.~\ref{bz} as a green hexagon.\\
\indent The nearest neighbour parametrized tight binding expression for band structure of strained graphene is given by Eq.~\eqref{eqEk}. We obtain the hopping parameters $t_1^{i}$, $t_2^{i}$ and $t_3^{i}$ for strained graphene from Ribeiro et al.(see Apendix~\ref{app1}) \cite{Ribeiro2009}. These parameters are extracted by fitting the band-structure obtained from ab initio calculations with Eq.~\eqref{eqEk}. Using band-structure inside the $1^{st}$ Brillouin zone, we compute mode density function at different strain values.
 \begin{equation}
E^{i}(k) = \pm \mid{t_1^{i}e^{-i\vec{k}\cdot{\vec{ a_1^{i}}}} + t_2^{i} + t_3^{i}e^{-i\vec{k}\cdot{\vec{ a_2^{i}}}} }\mid
\label{eqEk}
\end{equation}

\subsubsection{Mode density calculation} \label{subsec2}
\begin{figure}
	\subfigure[]{\includegraphics[height=0.2\textwidth,width=0.228\textwidth]{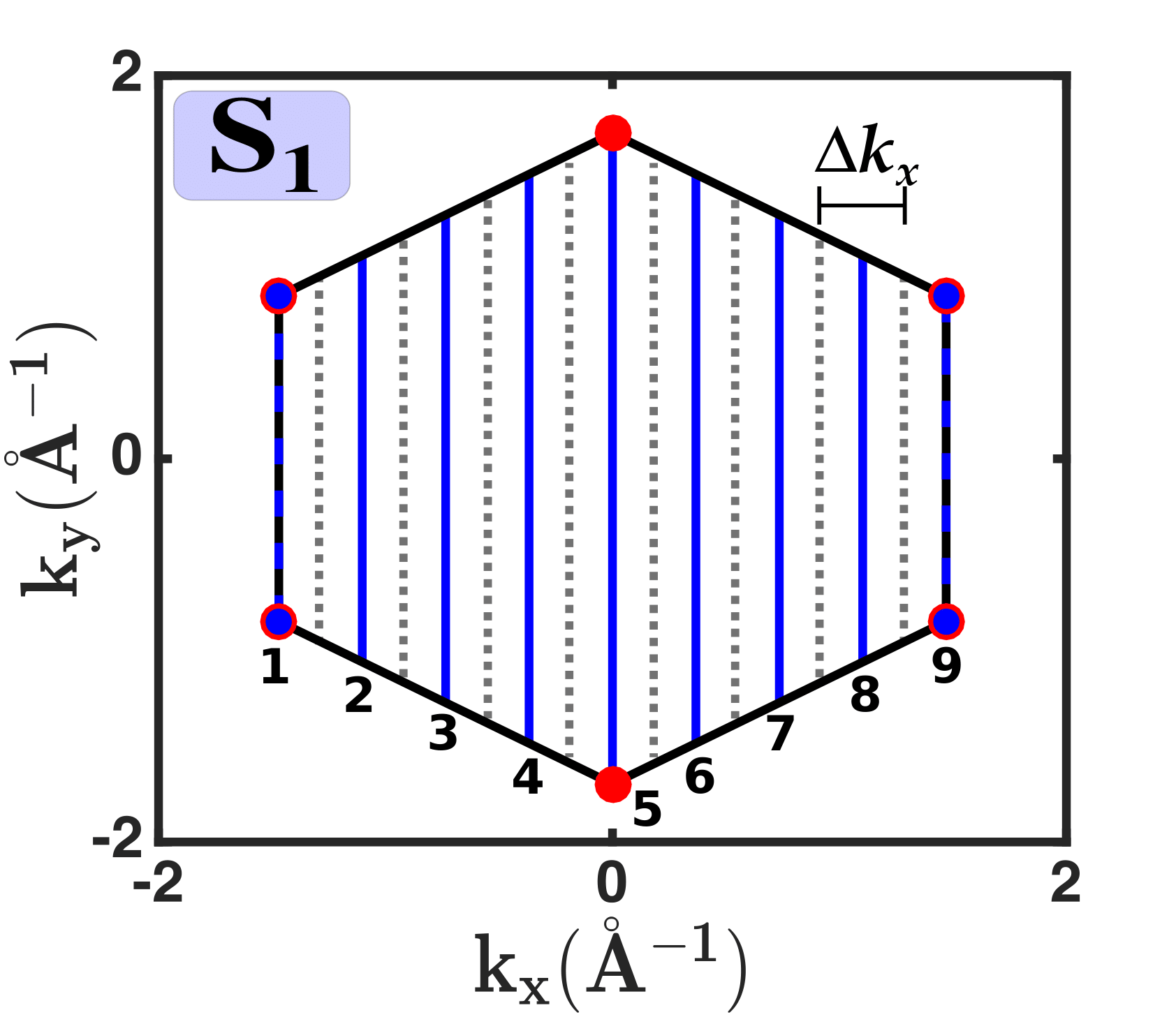}\label{Sub_S1}}
	\quad
	\subfigure[]{\includegraphics[height=0.2\textwidth,width=0.228\textwidth]{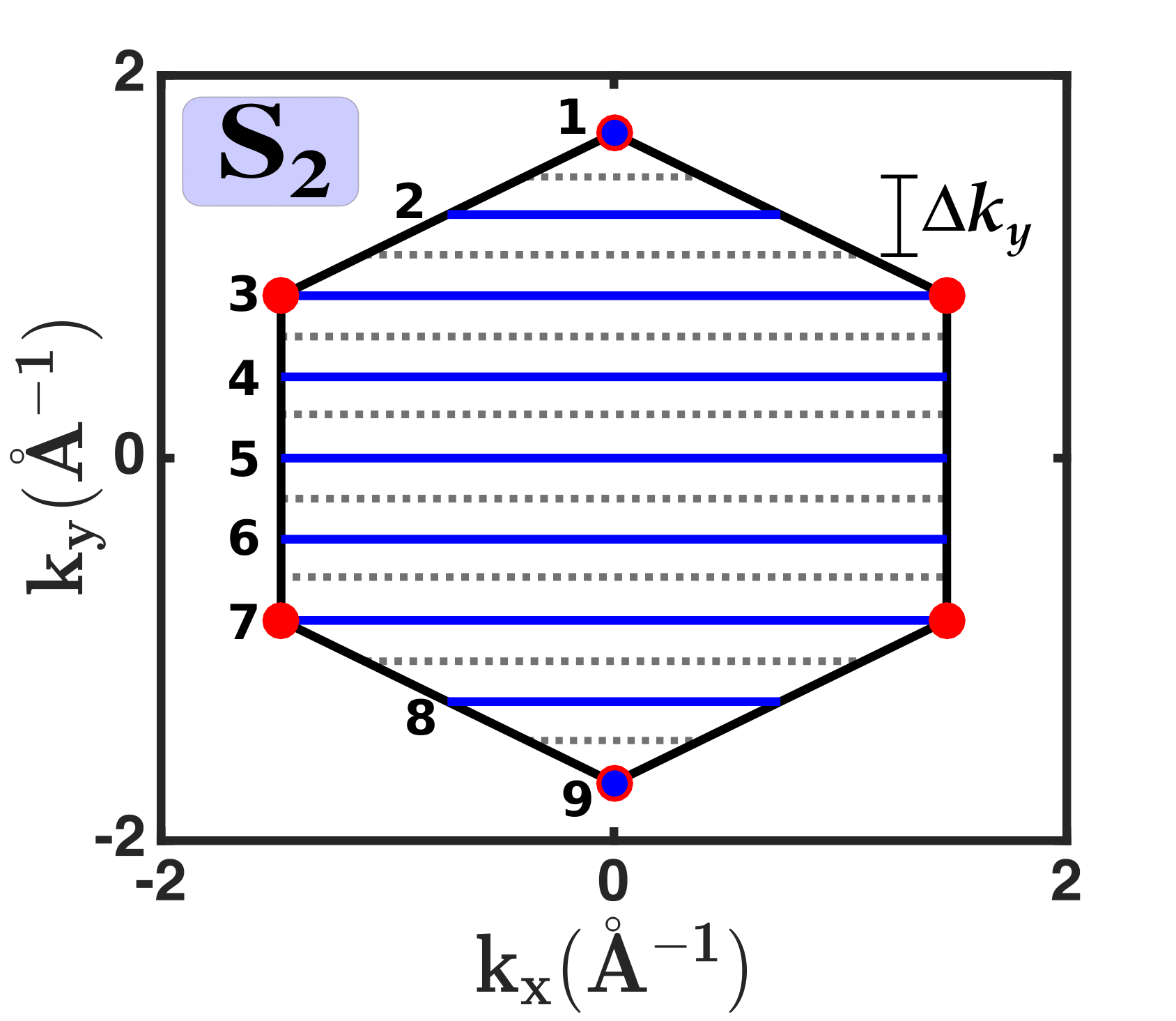}\label{Sub_S2}}	
	\quad
	\subfigure[]{\includegraphics[height=0.2\textwidth,width=0.228\textwidth]{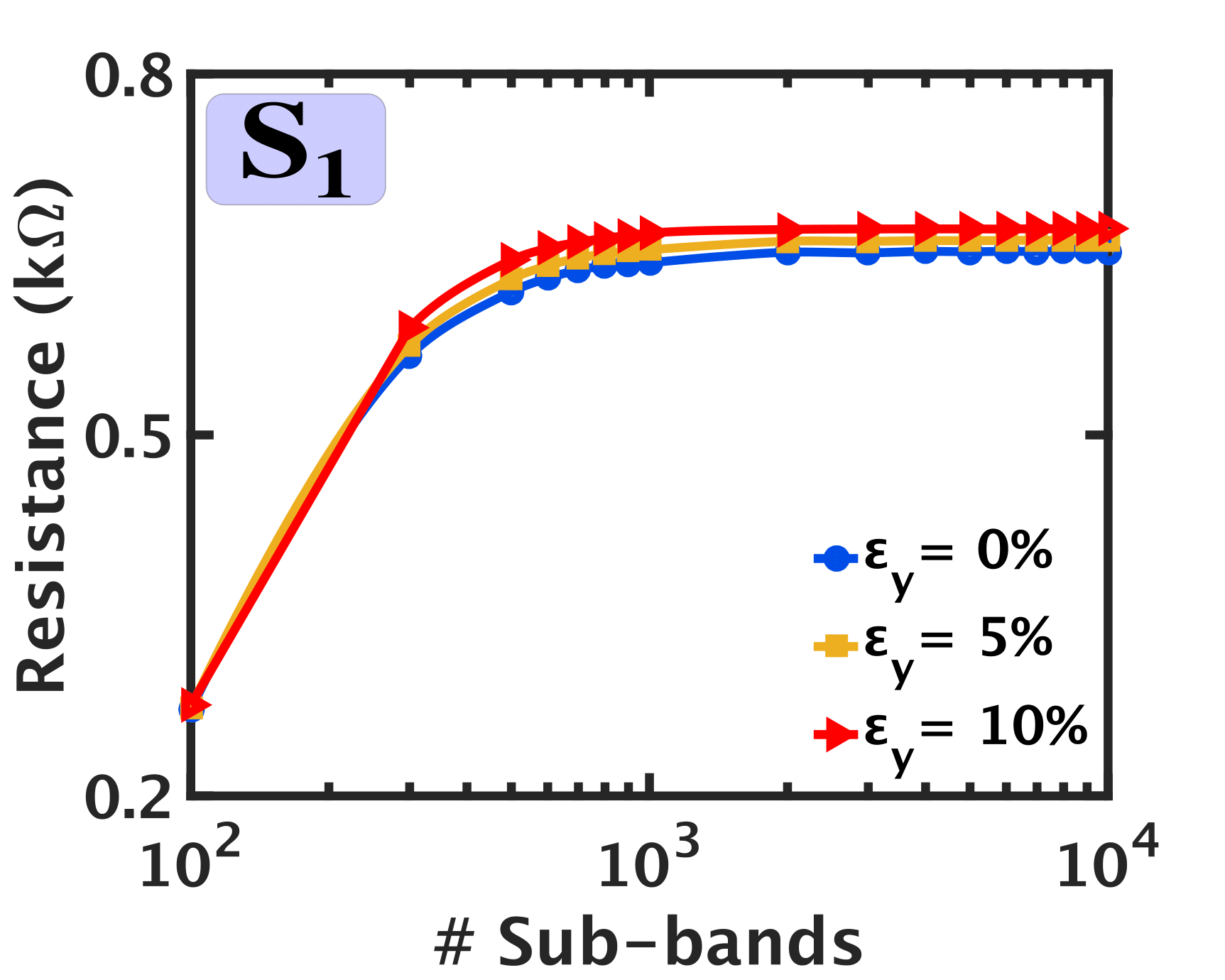}\label{R_S1}}
	\quad
	\subfigure[]{\includegraphics[height=0.2\textwidth,width=0.228\textwidth]{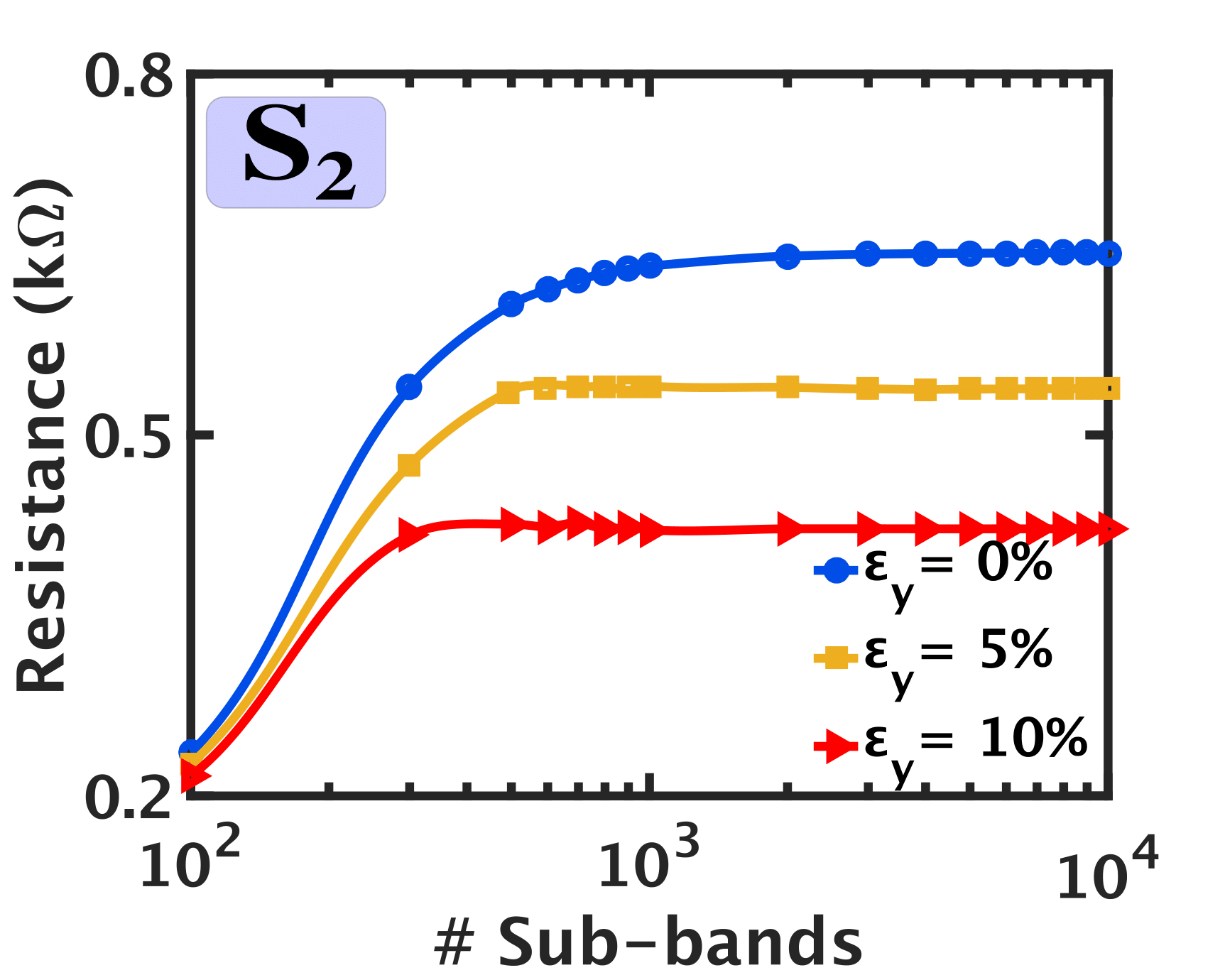}\label{R_S2}}

	\caption{ The path of subbands in the $1^{st}$ Brillouin zone of graphene in (a) setup $S_{1}$ and (b) setup $S_{2}$ for mode density calculation. Only nine segments are shown here for representational purpose. The variation of resistance with number of segments in (c) setup $S_{1}$, and in (d) setup $S_{2}$ for $0\%$, $5\%$ and $10\%$ strain. The value of resistance becomes constant above 1000 subbands in $S_{1}$ and $S_{2}$.}
			
	\label{sub-band}
\end{figure}

The most important step in GF calculation is determination of the mode density function. There are two ways to obtain mode density function in a ballistic conductor:
\begin{itemize}
  \item By counting the number of bands crossing a particular energy level (Band counting method).
  \item By using non equilibrium green's function method (NEGF).
\end{itemize}
Amongst these two methods,the band counting method is a relatively simpler technique for mode density calculation, provided dispersion relation is known. Here, we discuss the calculation of mode density function of graphene strained along armchair and zigzag directions from their band-structure. To calculate the mode density function of a graphene sheet, we must calculate the mode density of each transverse modes (TMs).\\ 
\indent TMs are formed due to quantum confinement along the transverse direction, leading to the quantization of momentum. Each point in the energy dispersion of a TM acts as a channel for electron transport. By counting the number of bands crossing a particular energy, we evaluate the mode density of a particular TM. The total mode density function is obtained by summing up the mode density of each TMs.\\
\indent For a 2D material like graphene, TMs are densely packed in $k_{x}$--$k_{y}$ plane. Thereby making the mode density evaluation, a difficult task. Thus, we compute the mode density function of graphene using a numerical technique that implements the band counting method. A similar method has been utilized to study the transport properties of germanium~\cite{Jeong2009}.\\
\indent The separation between the TMs in setup $S_{1}$ and setup $S_{2}$ is $2\pi/L^{i}_{cs}$, where $L^{i}_{cs}$ is the cross-sectional length at strain (`i\%').\\
\indent We divide the $1^{st}$ Brillouin zone of graphene in `$j$' equal segments along the transverse direction. Each segment contains a sub-band, as shown in Fig.~\ref{Sub_S1} and Fig.~\ref{Sub_S2}. We assume that the TMs that exists inside each segment containing the subband (say `$k_{\perp}$' shown with blue line), have the same mode density. The width of each segment containing the subband ($k_{\perp}$) is $\Delta k_{x}$ in $S_{1}$ and $\Delta k_{y}$ in $S_{2}$ as shown in Fig.~\ref{Sub_S1} and Fig.~\ref{Sub_S2} respectively. We gradually vary `$j$' from $10^2$ to $10^4$ and obtain the resistance variation of a unit micron wide graphene sheet. The plot of resistance versus the number of segments for $S_{1}$ is shown in Fig.~\ref{R_S1} and for $S_{2}$ is shown in Fig.~\ref{R_S2}. The value of resistance becomes constant above 1000 segments in both setups. The values of transmission, current density, and resistance obtained above 1000 segments are the actual values for the graphene sheet. The mode density of sub-band $(k_{\perp})$ assuming `p' energy minima and `q' energy maxima are present is given by:
\begin{equation}
M^{i}_{k_\perp}(E) = \sum_{p=1}^{p} \Theta(E\mp E^{i}_{p}) - \sum_{q=1}^{q} \Theta(E\mp E^{i}_{q}) 
\label{eqM}
\end{equation}
The negative sign in Eq.~\eqref{eqM} is used for conduction band whereas the positive sign is used for valence band. The detail derivation of Eq.~\eqref{eqM} is given in Appendix~\ref{app2}.\\ 
\indent Figure~\ref{sub-band} shows the path for 9 subbands. The collective mode density (per unit cross-sectional length) of all the TMs in the region $\Delta k_{x}$ (Fig.~\ref{Sub_S1}) or $\Delta k_{y}$ (Fig.~\ref{Sub_S2}), containing the subband `$k_{\perp}$', is given by:
\begin{equation}
 T^{i}_{k_{\perp}}(E) = P^{i}_{k_{\perp}}*M^{i}_{k_{\perp}}(E)
 \label{eqT1}
\end{equation}
where, $P^{i}_{k_{\perp}}$ is the prefactor and $M^{i}_{k_{\perp}}$ is the mode density of a subband `$k_{\perp}$'. For the edge subbands (with $k_{\perp}=$ 1 and $j$) in Fig.~\ref{sub-band}, the prefactors are $P^{i}_{1}$ and $P^{i}_{j}$, and for all the intermediate subbands (with $k_{\perp}=2$ and $j-1$), the prefactor is $P^{i}_{k_{\perp}}$ where $P^{i}_{1}$ = $P^{i}_{j}$ = $P^{i}_{k_{\perp}}/2$ and $P^{i}_{k_{\perp}}$= $ \Delta k/(2\pi)$ (refer to Appendix~\ref{app2}).

The total mode density or transmission per unit cross-sectional length of graphene is given by(see Appendix~\ref{app2}):
\begin{equation}
T^{i}(E)=P^{i}_{1}M^{i}_{1}(E)+\sum_{k_{\perp}=2}^{j-1} P_{k_{\perp}}^i M^{i}_{k_\perp}(E)+P_{j}^{i}M^{i}_{j}(E)
\label{eqT}
\end{equation}
\subsubsection{Evaluation of Gauge factor(GF)}
The equation for current density $J^{i}(V)$ using Landauer formula \cite{supriyo2012lessons} is given by:
\begin{equation}
J^{i}(V)= \frac{q}{h} \int_{-\infty}^{\infty} T^i(E)[f_{1}(E-{\mu_{1}})-f_{2}(E-{\mu_{2}})]dE
\label{eqJ}	
\end{equation}
 We evaluate the current-density in linear regime i.e. a few $kT$ near the Fermi energy. Here, $kT$ is thermal energy at room temperature. The quantity $r^{i}$ is given by:
\begin{equation}
r^{i} = \dfrac{dV}{dJ^{i}_{j}}
\label{eqr}
\end{equation}
$r^{i}$ is resistance of a unit cross-sectional length of graphene. The value of resistance ($R^{i}_{j}$) is given by:
\begin{equation}
R^{i} = \frac{r^{i}}{L_{cs}(1+\nu i)}
\label{eqR}
\end{equation}
 where, $L_{cs}$ is cross sectional length of graphene at $0 \%$ strain, $\nu$ is poisson's ratio and $i$ is the magnitude of uniaxial strain $\varepsilon_{x}$ or $\varepsilon_{y}$. Finally, the gauge factor is given by:
 \begin{equation}
 GF= \frac{ R^{i}-R^{0}}{{R^{0}}\varepsilon}= \frac{1}{\varepsilon}\Bigg[{\frac {r^{i}}{r^{0} (1+\nu \varepsilon)} -1\Bigg]}
 \label{eqGF}
 \end{equation}
 where, $R^0$ is the value of resistance at zero strain and $R^{i}$ is the resistance at $i\%$ strain. 
 \begin{figure*}[]
	\subfigure[]{\includegraphics[height=0.215\textwidth,width=0.225\textwidth]{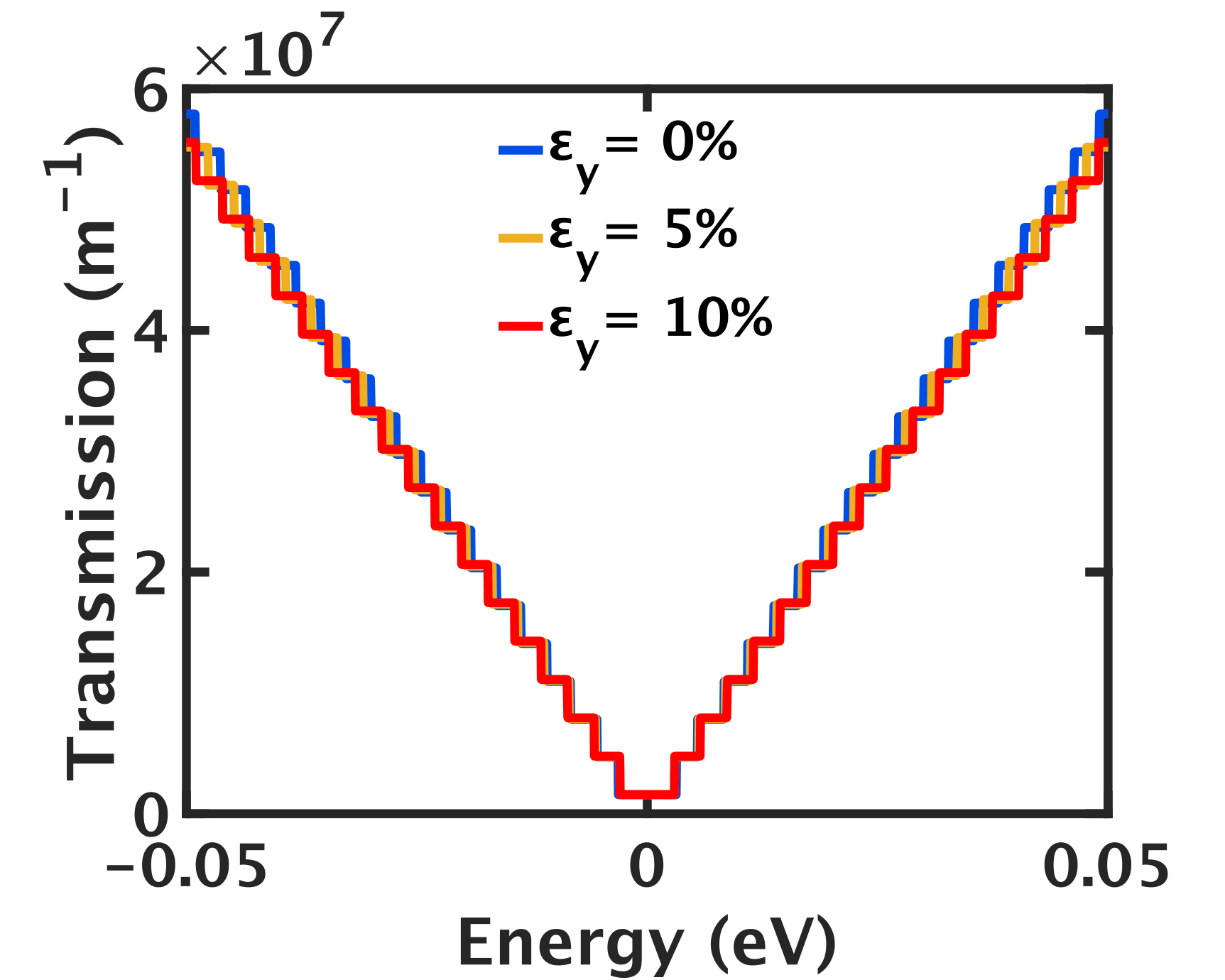}\label{TE_ZZ_LS}}
	\quad
	\subfigure[]{\includegraphics[height=0.21\textwidth,width=0.229\textwidth]{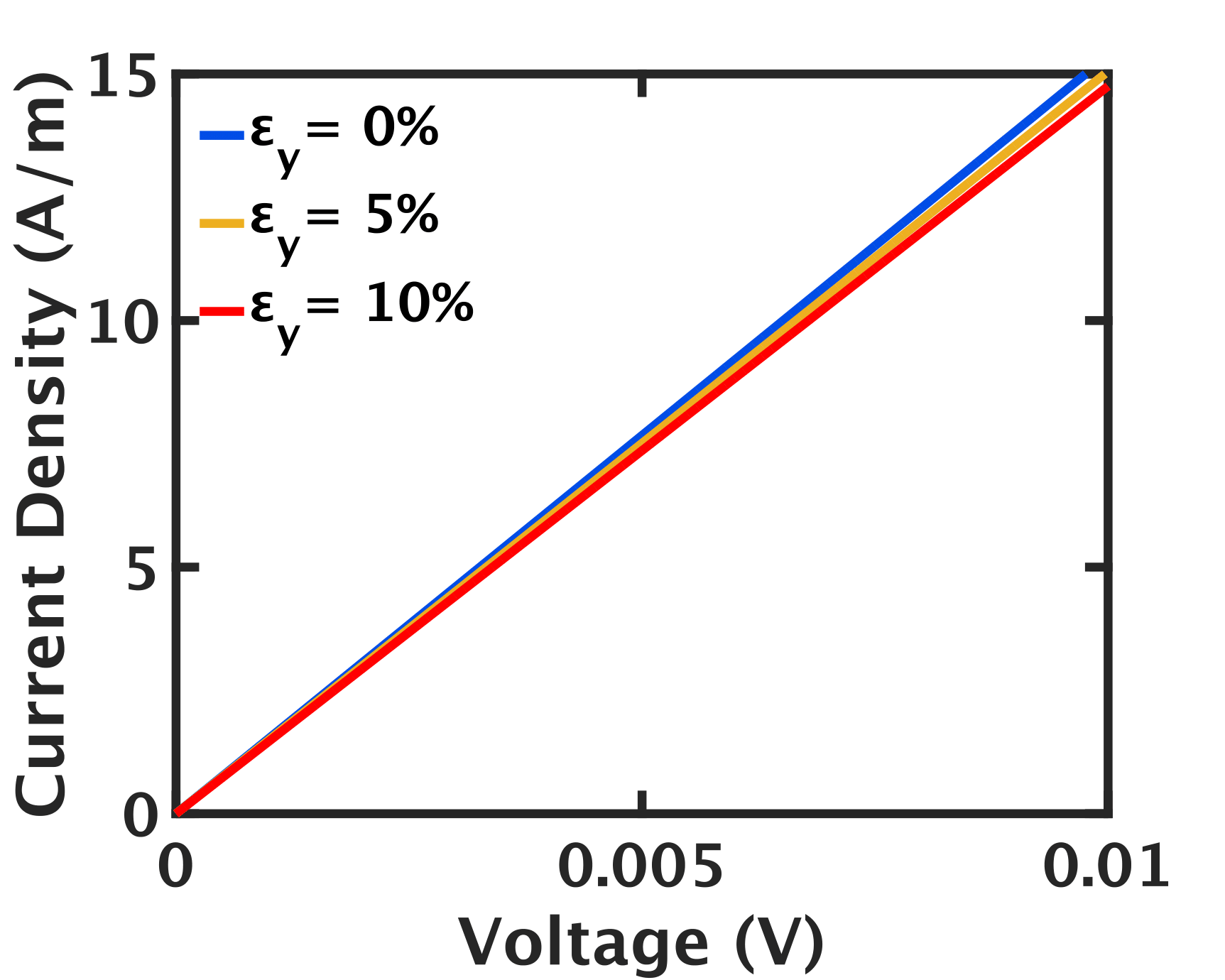}\label{IV_ZZ_LS}}
	\quad
	\subfigure[]{\includegraphics[height=0.21\textwidth,width=0.229\textwidth]{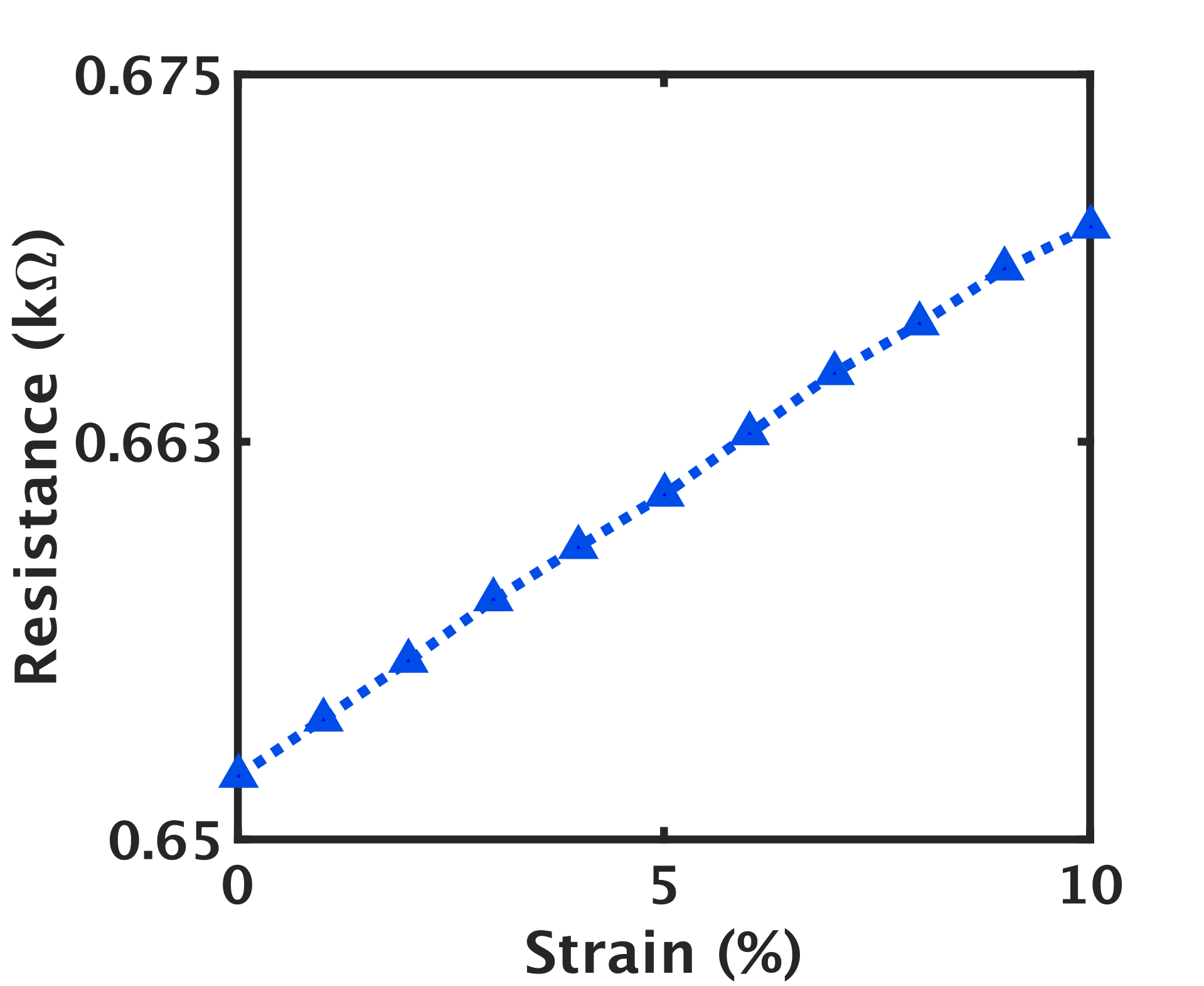}\label{R_ZZ_LS}}
	\quad
	\subfigure[]{\includegraphics[height=0.21\textwidth,width=0.229\textwidth]{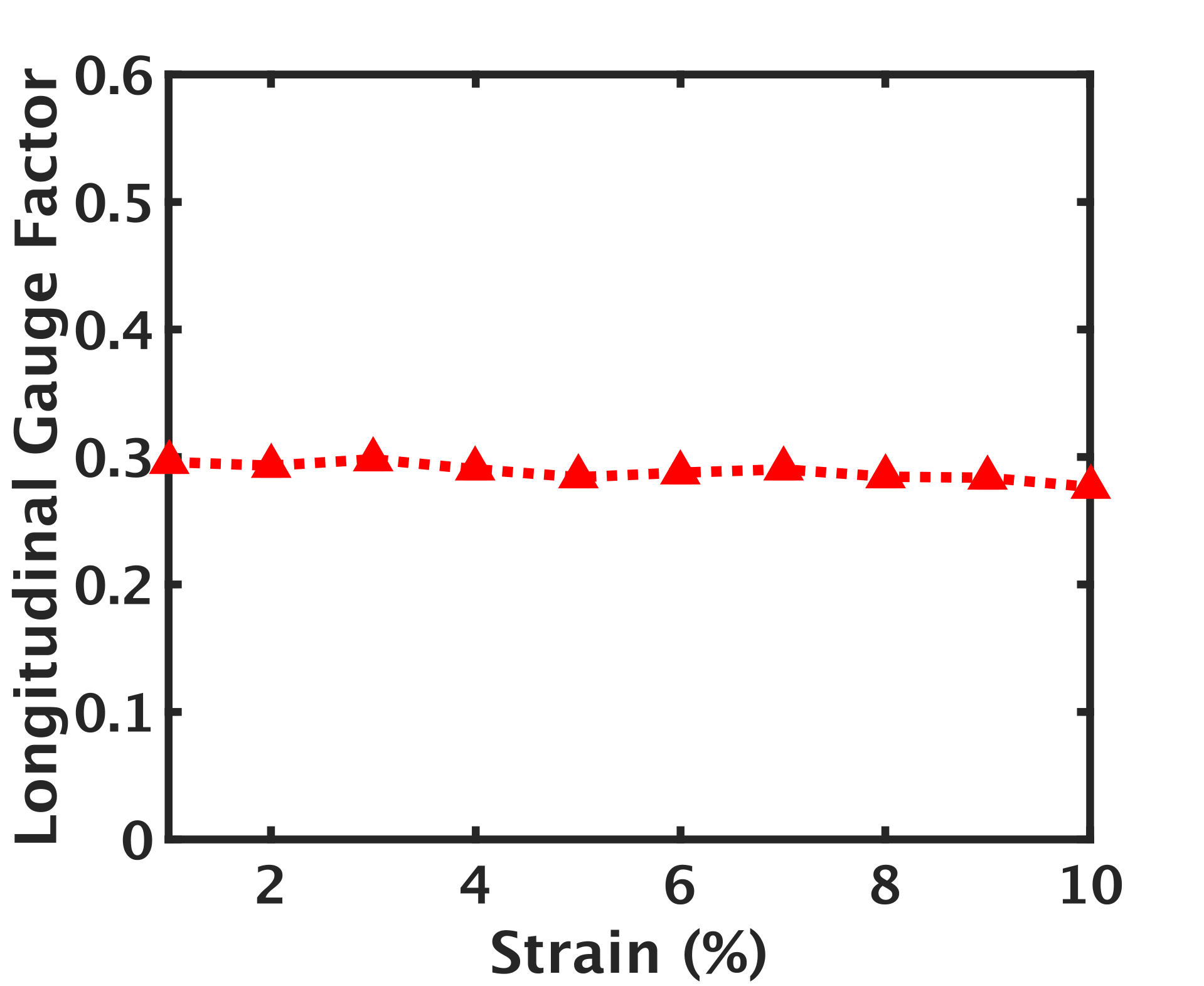}\label{GF_ZZ_LS}}

	\caption{ Strain($\varepsilon_{y}$) dependent transport properties in setup $S_{1}$ of graphene.(a) Plot of transmission versus energy for $0\%$, $5\%$ and $10\%$ strain. Transmission decreases with an increase in strain.(b) Plot of current density versus voltage at $0\%$, $5\%$ and $10\%$ strain. The magnitude of current density decreases with strain. (c) Plot of resistance versus strain of a $1\mu m$ wide graphene sheet. The resistance increases linearly with strain (d) Plot of longitudinal gauge factor versus strain. The average LGF is $\approx 0.3$.} 
	\label{Trans_prop_zz_LS}
\end{figure*}

\begin{figure*}[]
   	\subfigure[]{\includegraphics[height=0.215\textwidth,width=0.225\textwidth]{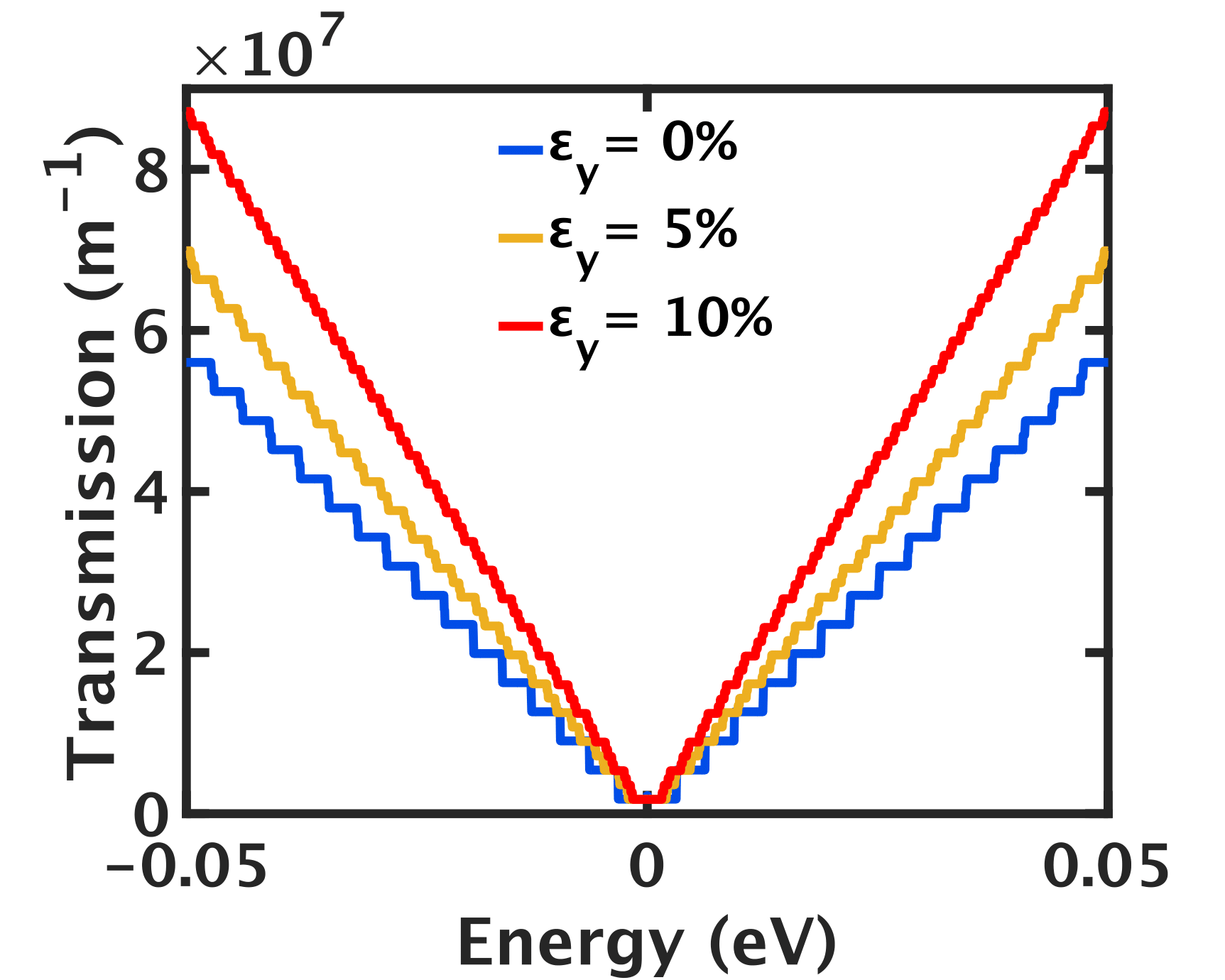}\label{TE_ZZ_TS}}
	\quad
	\subfigure[]{\includegraphics[height=0.21\textwidth,width=0.229\textwidth]{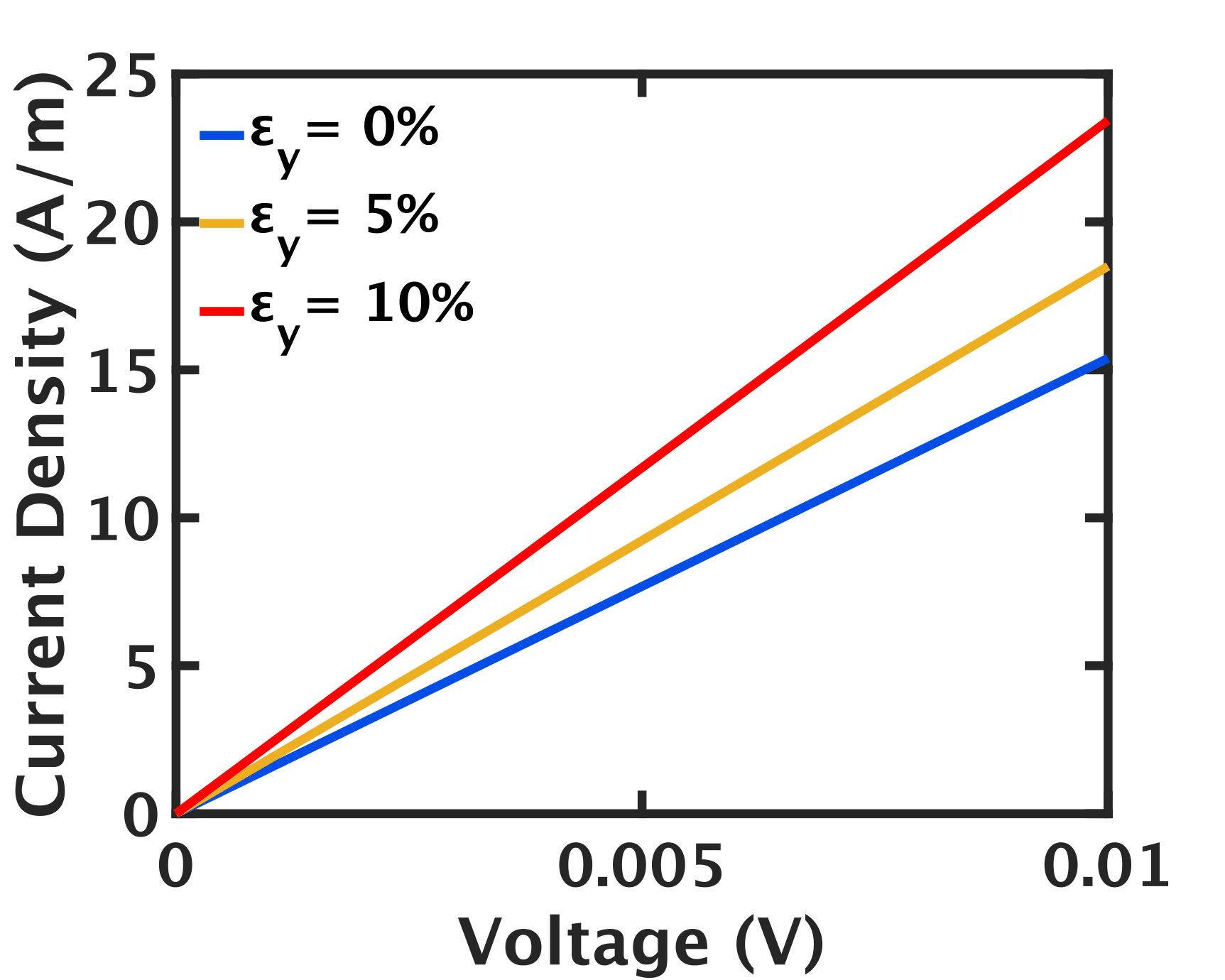}\label{IV_ZZ_TS}}
	\quad
	\subfigure[]{\includegraphics[height=0.21\textwidth,width=0.229\textwidth]{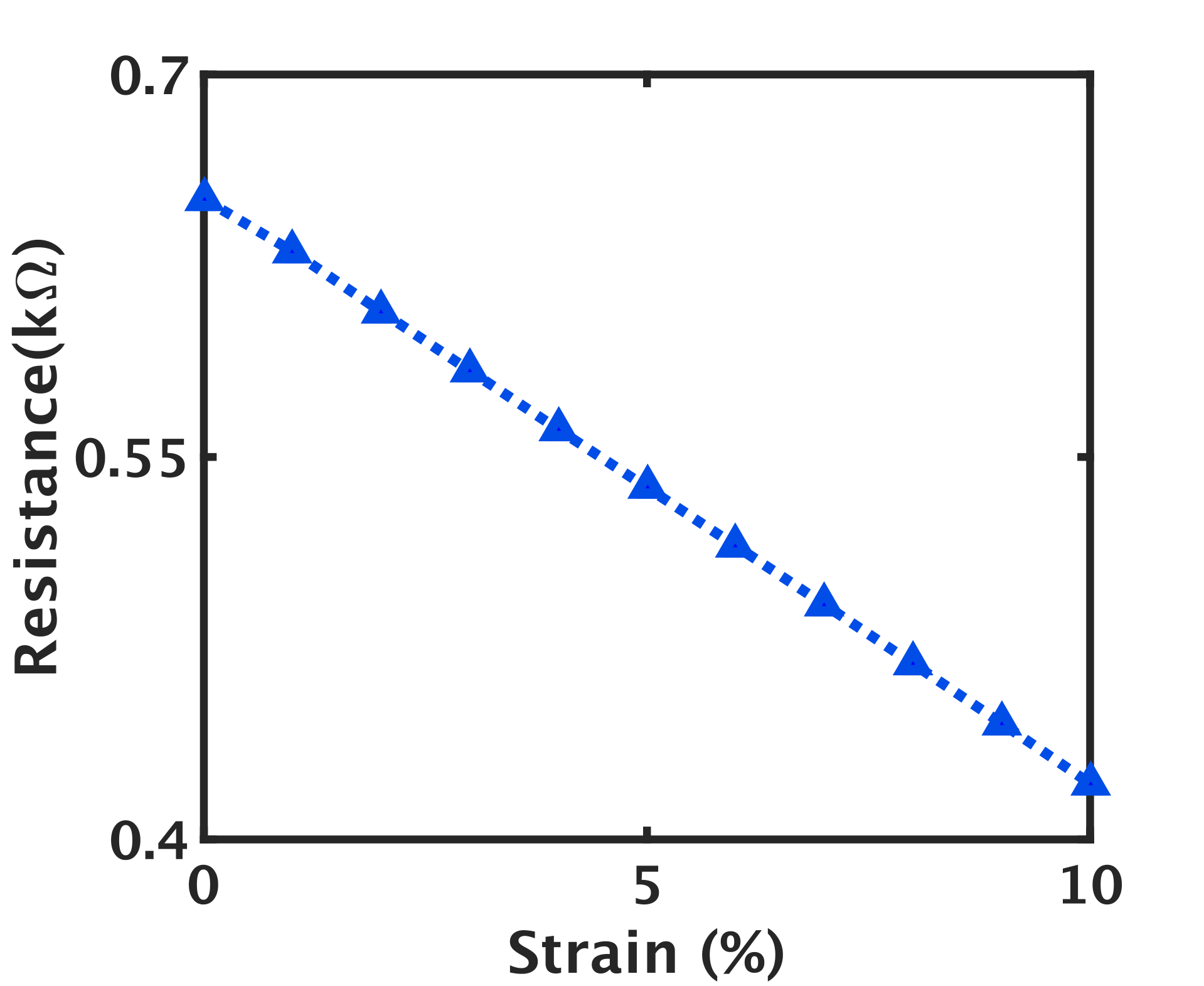}\label{R_ZZ_TS}}
	\quad
	\subfigure[]{\includegraphics[height=0.21\textwidth,width=0.229\textwidth]{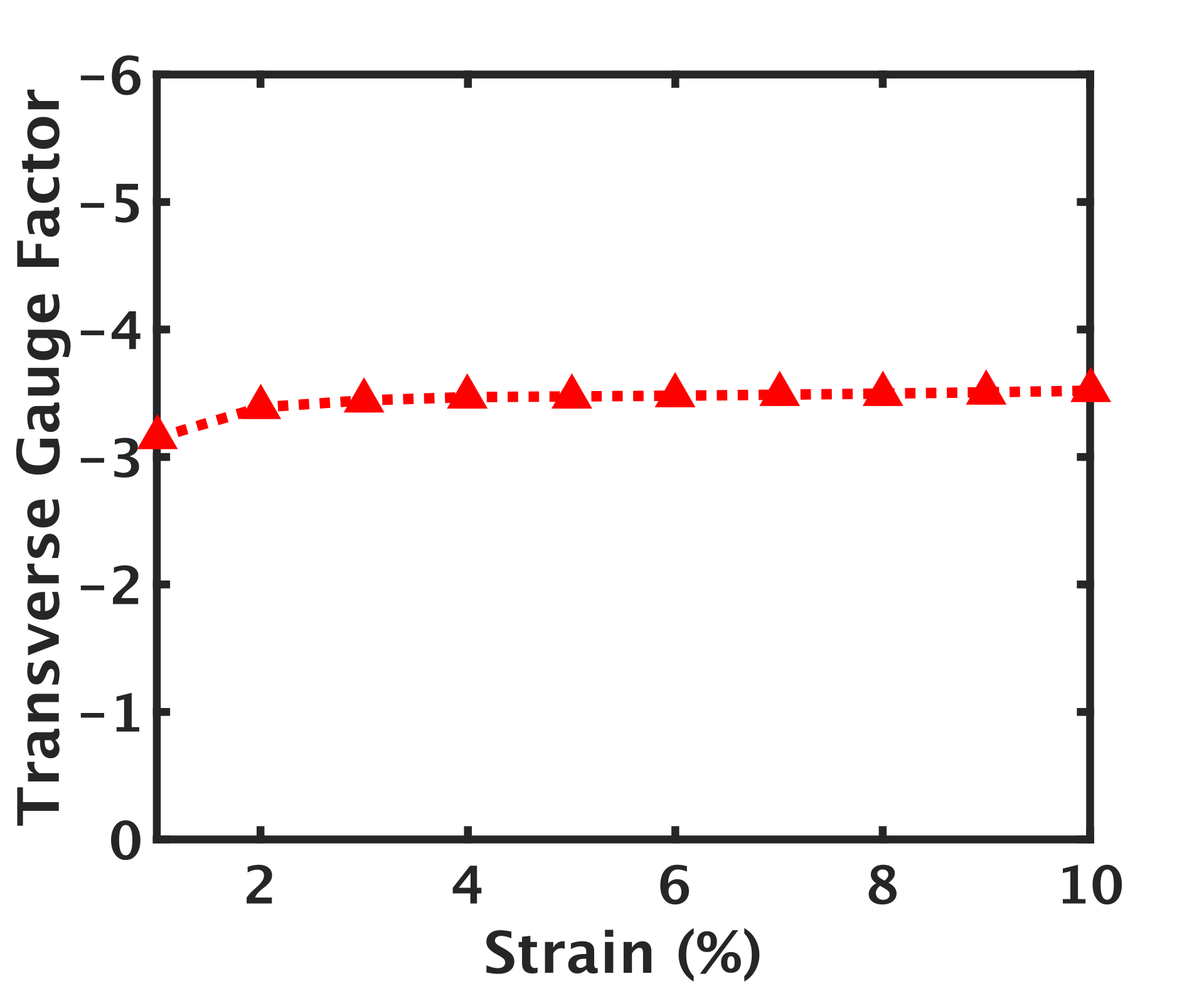}\label{GF_ZZ_TS}}
	
	\caption{Strain($\varepsilon_{y}$) dependent transport properties in setup $S_{2}$ of graphene.(a) Plot of transmission versus energy for $0\%$, $5\%$ and $10\%$ strain. The transmission increases with an increase in strain.(b) Plot of current density versus voltage at $0\%$, $5\%$ and $10\%$ strain. The magnitude of current density increases with strain. (c) Plot of transverse resistance versus strain of a $1\mu m$ wide graphene sheet. The resistance decreases linearly with strain (d) Plot of transverse gauge factor versus strain. The average TGF is $\approx -3.3$.}  
   \label{Trans_prop_zz_TS}
\end{figure*}

\section{Results and Discussions } \label{section3}

\begin{figure}
	\centering
	\subfigure[]{\includegraphics[height=0.2\textwidth,width=0.22\textwidth]{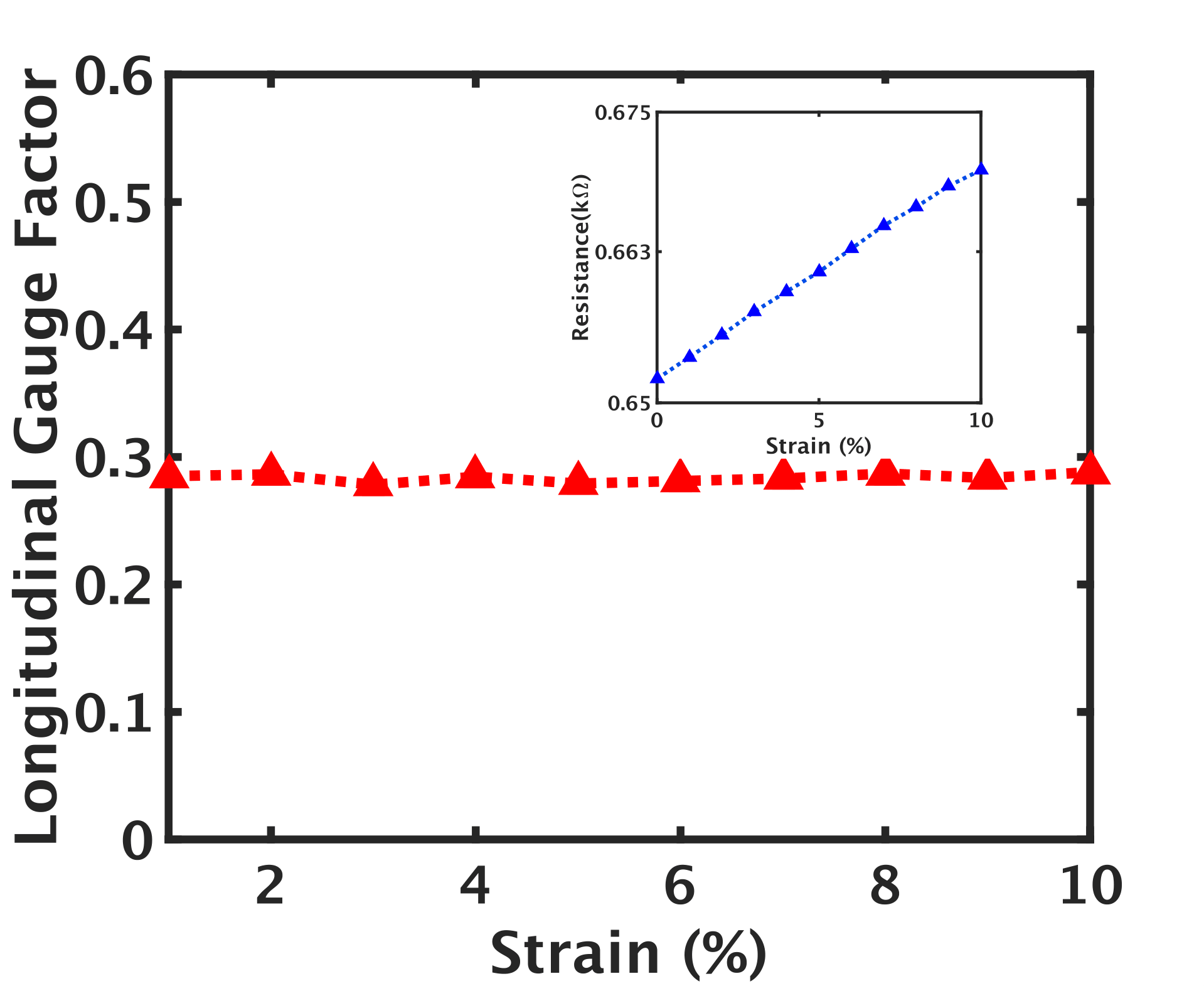}\label{GF_AC_LS}}
	\quad
	\subfigure[]{\includegraphics[height=0.2\textwidth,width=0.22\textwidth]{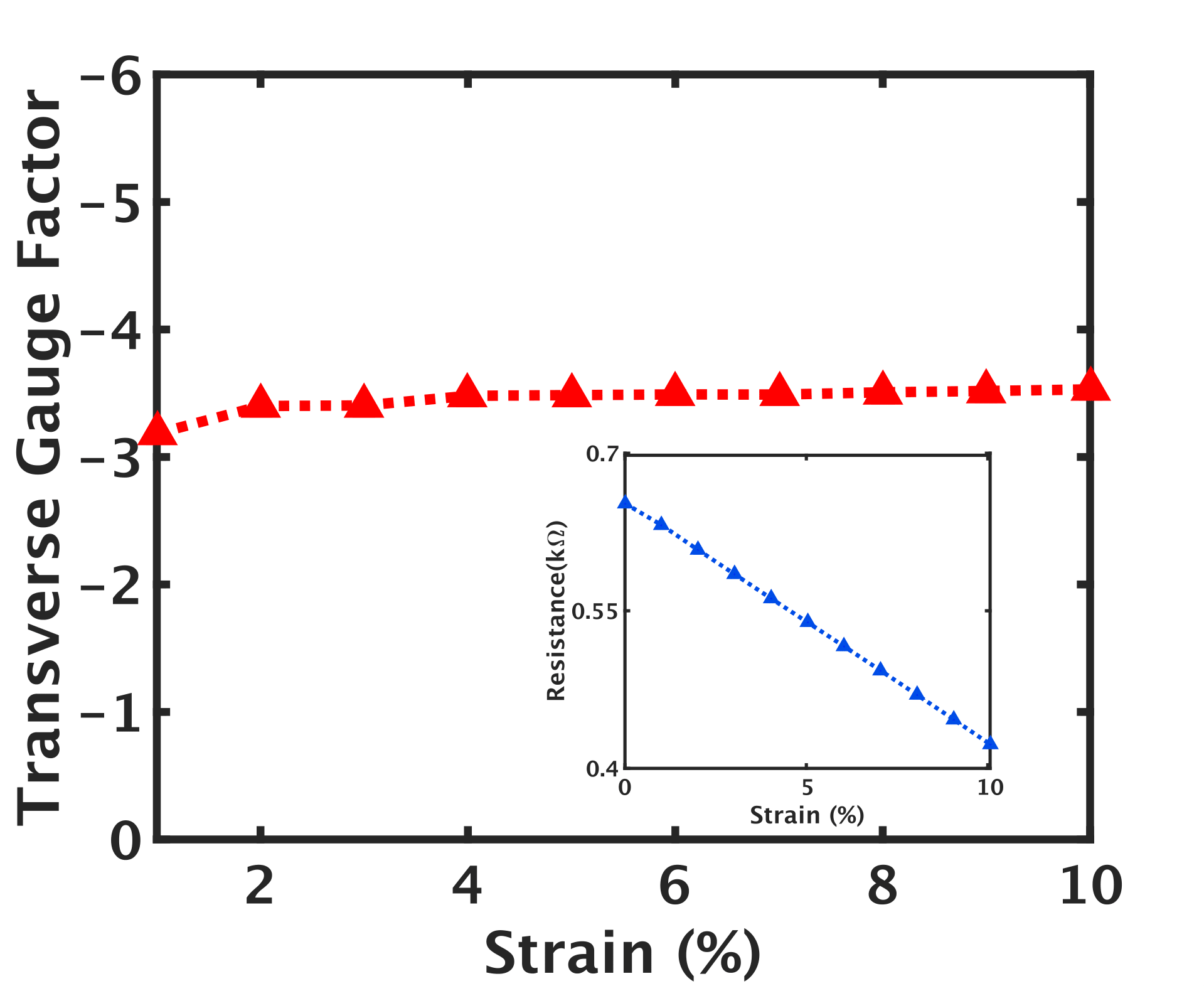}\label{GF_AC_TS}}
	
	\caption{(a) The longitudinal gauge factor of graphene for strain along the armchair direction. (b) The transverse gauge factor of graphene for strain along the armchair direction. The plot of resistance versus strain in shown as inset figures in (a) and (b). The longitudinal and transverse gauge factors of armchair graphene is same as zigzag graphene.}
	
	\label{GF_AC}
\end{figure}

We discuss the strain-dependent transport properties along zigzag direction in the ballistic regime. We further evaluate the longitudinal and transverse gauge factor along the zigzag direction and then compare it with that of the armchair direction. The piezoresistance effect is due to the distortion of Dirac cones and change in separation of TMs with applied strain. The transport properties of $S_{1}$ are shown in Fig.~\ref{Trans_prop_zz_LS} and $S_{2}$ are shown in Fig.~\ref{Trans_prop_zz_TS}.\\    
\indent The transmission increases with energy at a particular strain, as shown in Fig.~\ref{TE_ZZ_LS} and Fig.~\ref{TE_ZZ_TS}. It can be inferred from Fig.~\ref{TE_ZZ_LS} that in setup $S_1$ with an applied strain, the transmission decreases due to a decrease in the mode density as explained in the subsequent subsections. This reduction in transmission with increase in strain leads to a reduction in current density (Fig.~\ref{IV_ZZ_LS}) and finally an increase in the resistance (Fig.~\ref{R_ZZ_LS}). The average LGF of $S_{1}$ is 0.3 (Fig.~\ref{GF_ZZ_LS}). Whereas, in setup $S_2$, the transmission increases significantly with  applied strain as shown in Fig.~\ref{TE_ZZ_TS}. As a result, the current density increases substantially in Fig.\ref{IV_ZZ_TS} and finally the resistance decreases in Fig.\ref{R_ZZ_TS}. The average TGF of $S_{2}$ is -3.3 (Fig.~\ref{GF_ZZ_TS}).\\
\indent Our calculated value of resistivity of suspended graphene at $0\%$ strain is consistent with earlier work on suspended graphene by Adam et al. \cite{Das2008}. Here, we demonstrate a larger value of TGF  ($\approx$ 10 times) compared to LGF. Moreover, the longitudinal and transverse piezoresistance characteristics of armchair configuration is exactly identical to that of zigzag configuration (refer Fig.~\ref{GF_AC}).\\
\indent The linear variation of resistance in $S_{1}$ (Fig.~\ref{R_ZZ_LS}) and $S_{2}$ (Fig.~\ref{R_ZZ_TS}) with strain is especially useful in strain sensing. Our results suggest that suspended graphene based strain sensors can be easily calibrated to measure physical quantities such as pressure, force, tension, etc. at high strain. The $S_{1}$ setup is more sensitive than $S_{2}$ setup due to the higher value of gauge factor. High strain sensors are useful in structural health monitoring, stretchable electronics, etc. High strain sensors have been studied previously for zinc oxide nanowire based flexible films \cite{xiao2011} and graphene-rubber composite \cite{boland2014}. Due to very small thickness, graphene has high sensitivity per unit area for pressure sensing \cite{shih2001,Smith2016}. Therefore, TGF configuration can be easily calibrated with pressure to make very sensitive nano pressure sensors for high strain application. \\
\indent Physics behind the sizeable variation of longitudinal and transverse gauge factors is elaborately explained in subsequent subsections.

\begin{figure*}
\centering
	
	\subfigure[]{\includegraphics[height=0.3\textwidth,width=0.33\textwidth]{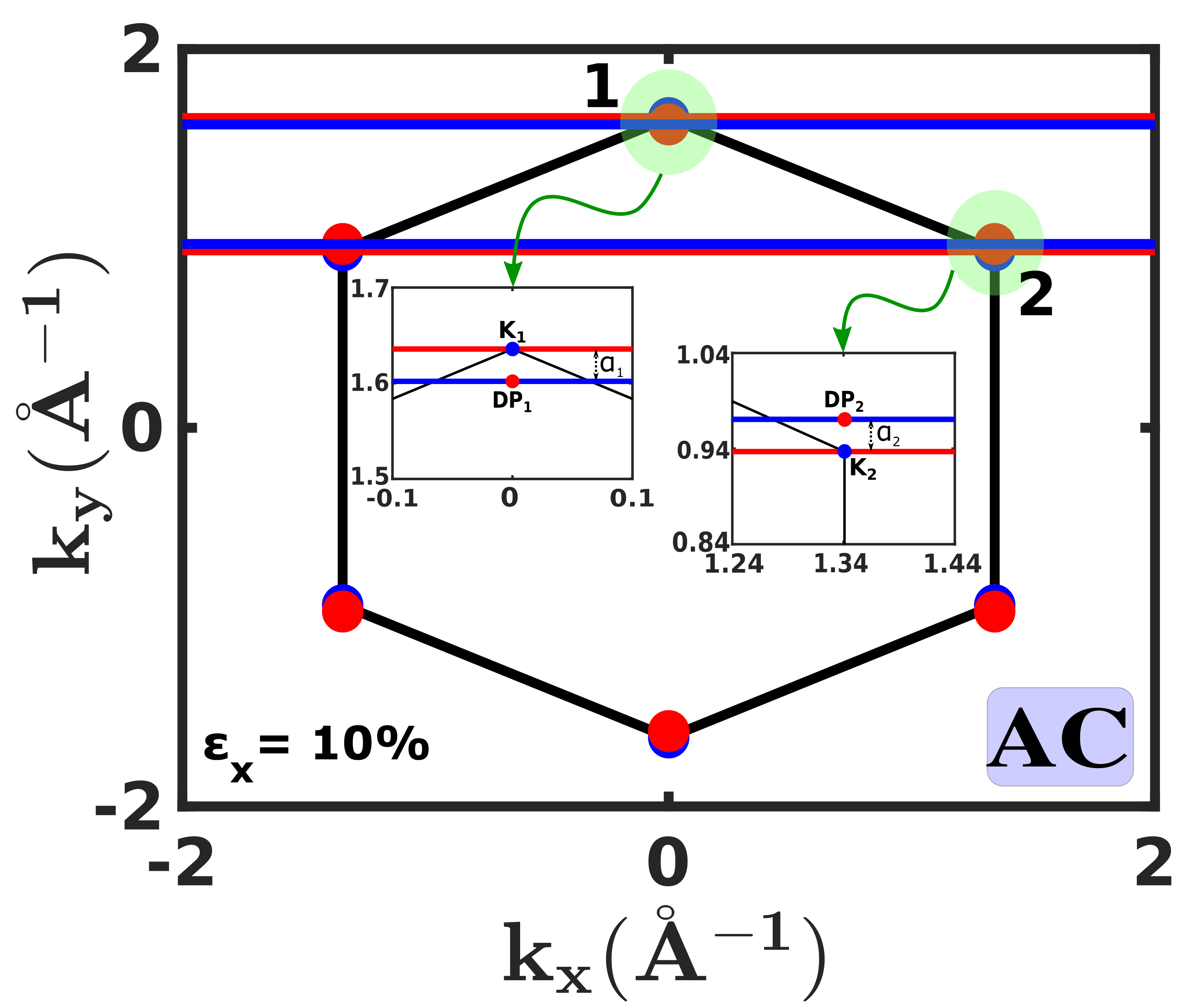}\label{ac_0_dp1_all}}
	\quad
	\subfigure[]{\includegraphics[height=0.3\textwidth,width=0.33\textwidth]{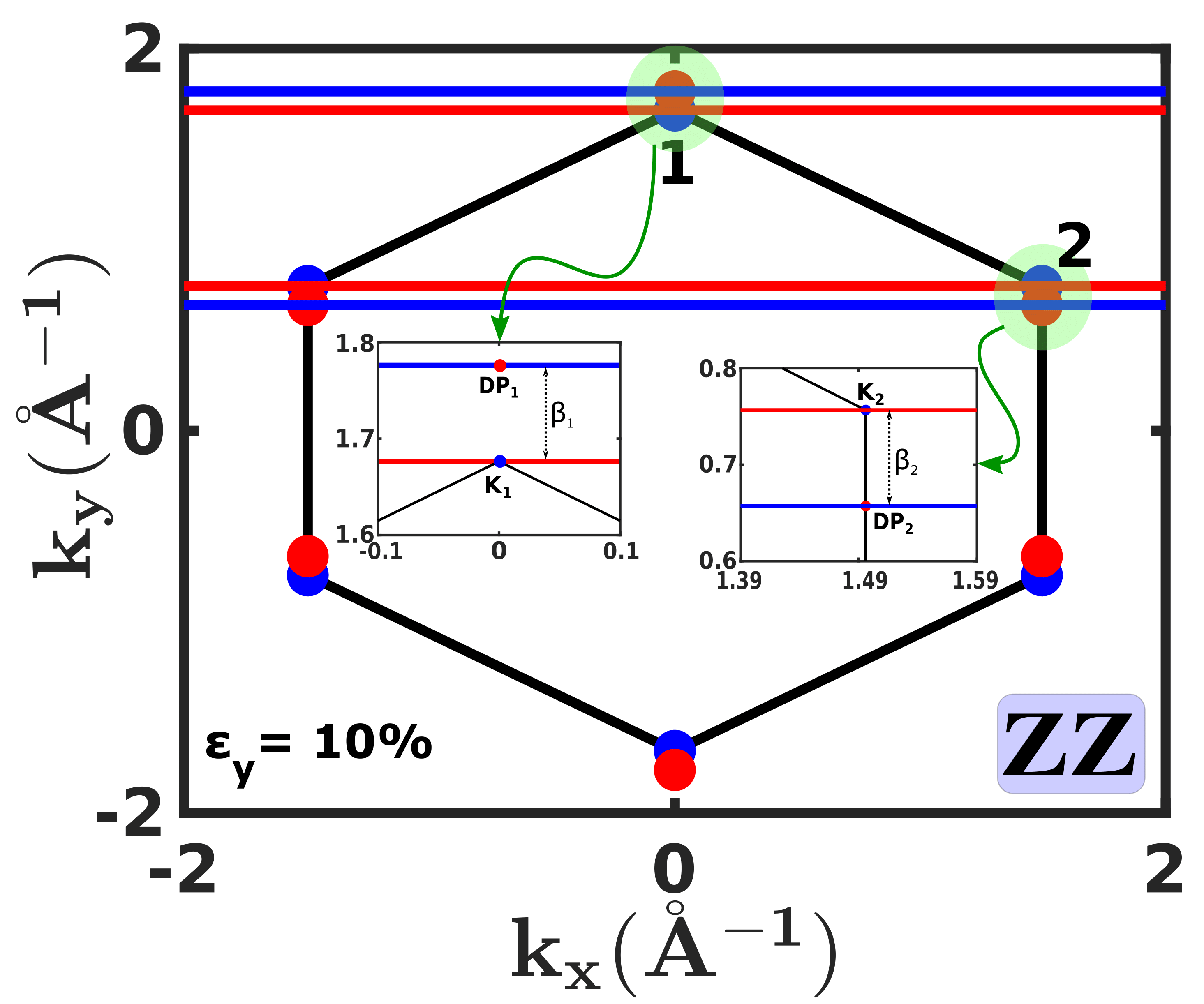}\label{zz_0_dp1_all}}
	
	\caption{(a) Shift in Dirac points $DP_{1}$ ($\alpha_{1}$) and $DP_{2}$ ($\alpha_{2}$) due to strain ($\varepsilon_{x}=10\%$) along armchair direction. $DP_{1}$ moves inside the Brillouin zone along the line joining $K_{1}$ and $K_{4}$ whereas $DP_{2}$ moves out of the Brilluoin zone along the line joining $K_{2}$ and $K_{3}$. (b)  Shift in Dirac points $DP_{1}$ ($\beta_{1}$) and $DP_{2}$ ($\beta_{2}$) due to strain ($\varepsilon_{y}=10\%$) along zigzag direction. $DP_{1}$ moves out of the Brillouin zone along the line joining $K_{1}$ and $K_{4}$ whereas $DP_{2}$ moves inside of the Brilluoin zone along the line joining $K_{2}$ and $K_{3}$. Inset figures in (a) and (b) shows the magnified view of the shift in Dirac points $DP_{1}$ and $DP_{2}$.}
	
	\label{dp_sub}
\end{figure*}

\begin{figure}
    \subfigure[]{\includegraphics[height=0.6\textwidth,width=0.495\textwidth]{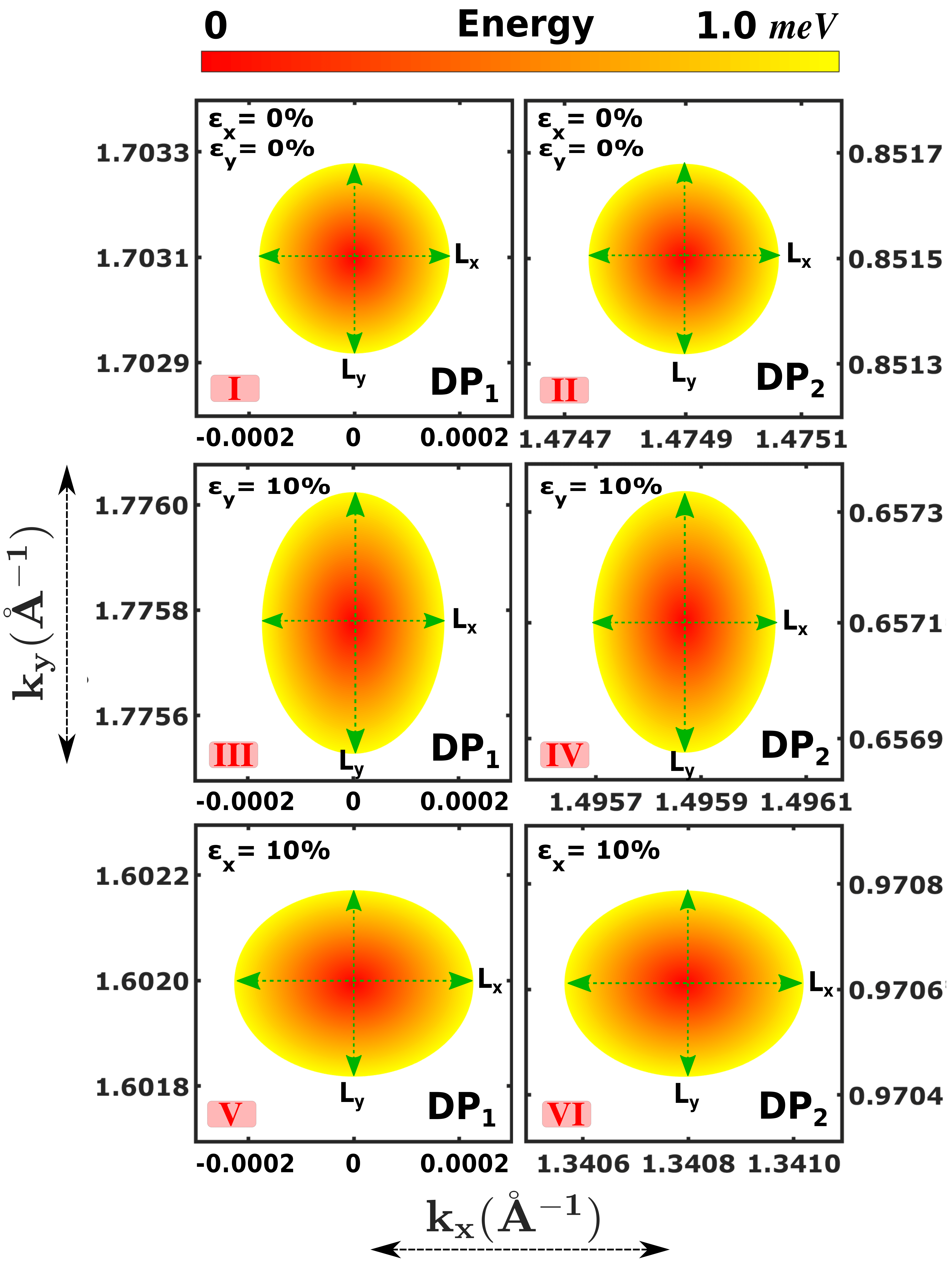}}

	\caption{Top view of the Dirac cone. I. $DP_{1}$ at $\varepsilon_{x}$=$\varepsilon_{y}$=$0\%$. II. $DP_{2}$ at $\varepsilon_{x}$=$\varepsilon_{y}$=$0\%$. III. $DP_{1}$ at $\varepsilon_{y}$=$10\%$. IV. $DP_{2}$ at $\varepsilon_{y}$=$10\%$. V. $DP_{1}$ at $\varepsilon_{x}$=$10\%$. VI. $DP_{1}$ at $\varepsilon_{x}$=$10\%$. The dimension of $DP_{1}$ and $DP_{2}$ are identical for the same strain. The major axis and minor axis of the dirac cones are equal for same magnitude of strain along x and y direction.  }
		
	\label{dc_strain}
\end{figure}

\begin{figure}

\subfigure[]{\includegraphics[height=0.40\textwidth,width=0.40\textwidth]{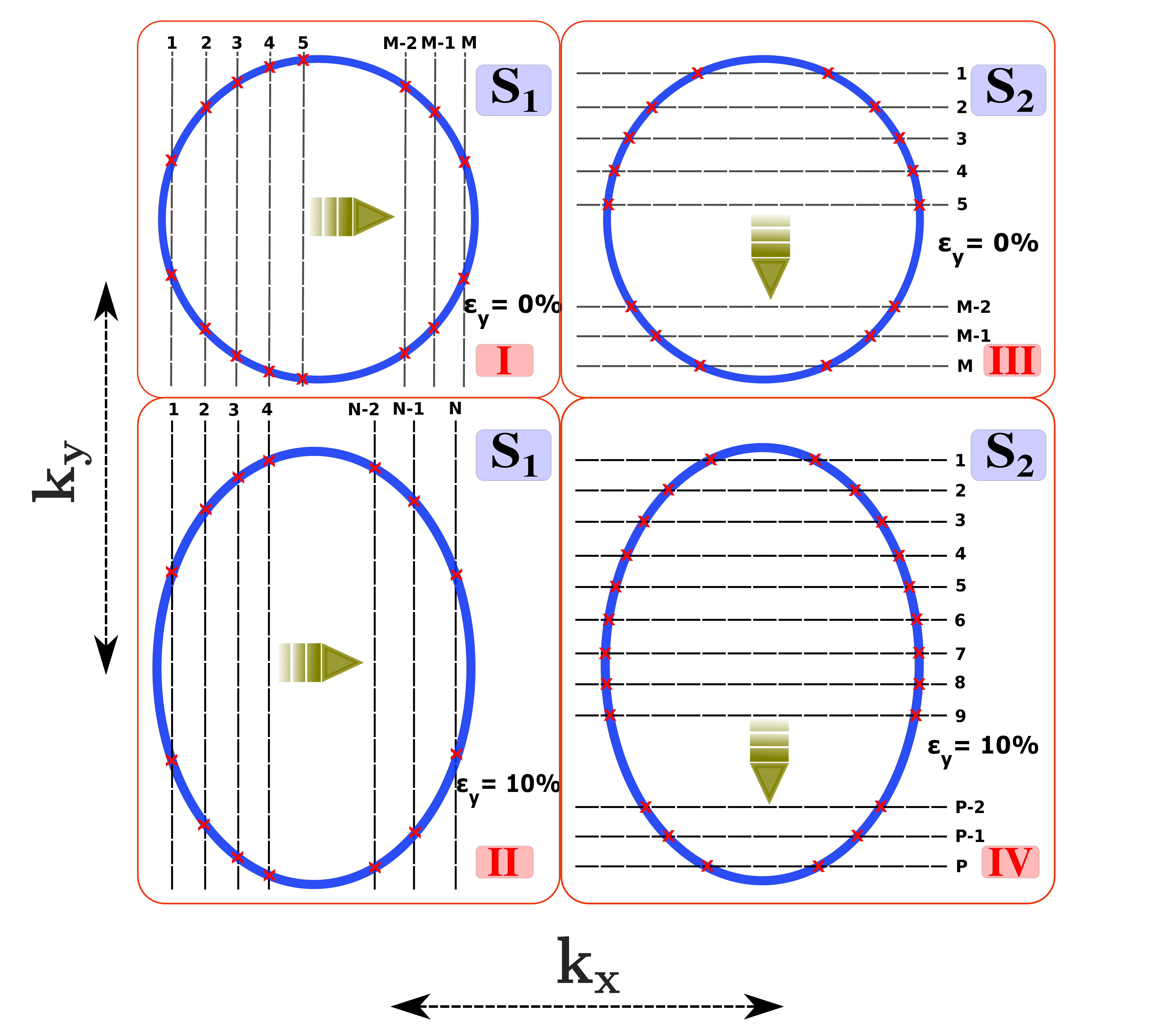}\label{modes}}

\subfigure[]{\includegraphics[height=0.30\textwidth,width=0.42\textwidth]{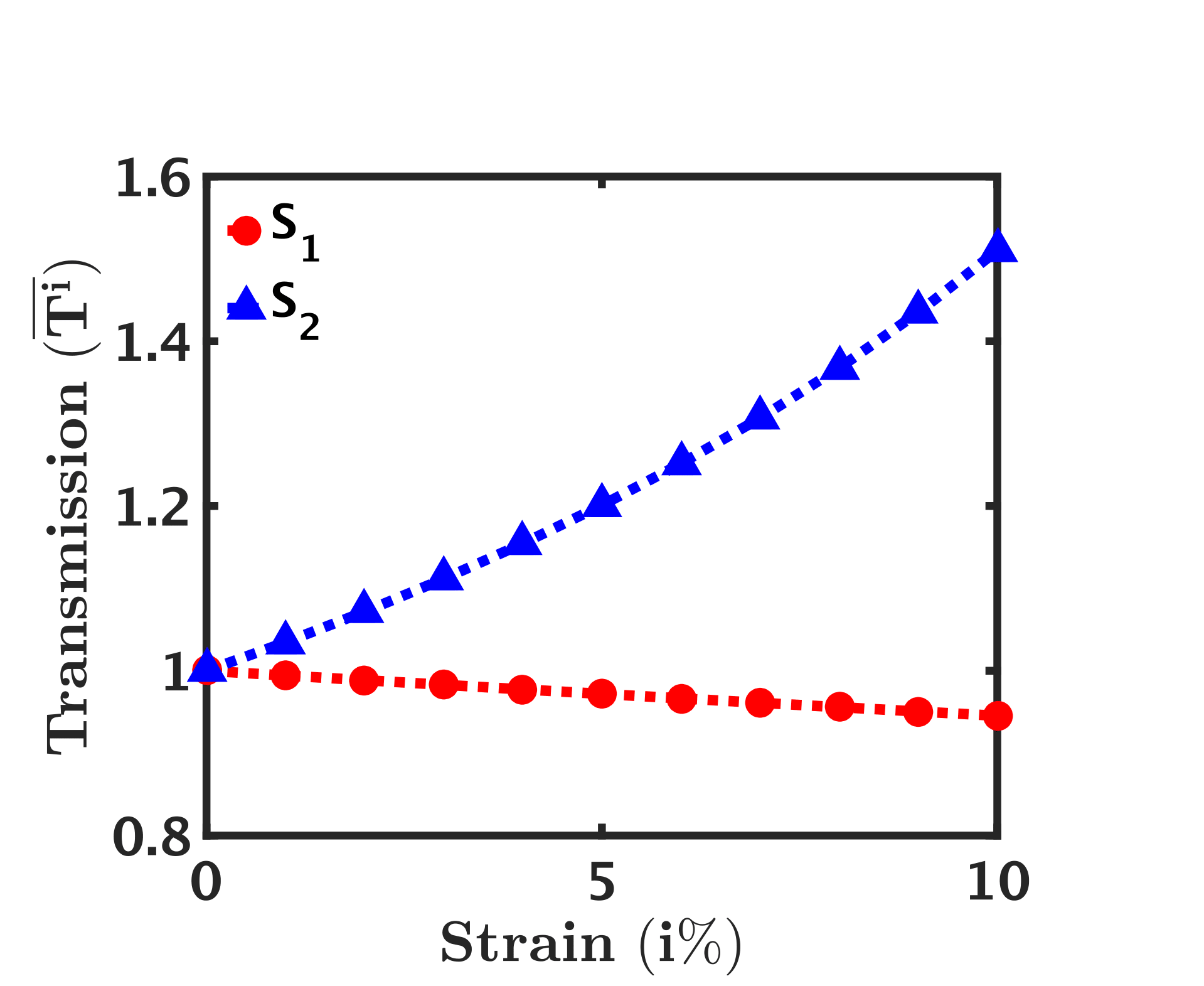}\label{norm_transmission}}
	
\caption {(a) Schematic diagram of the constant energy surface in Dirac cone and TMs at 1 meV energy. I) Dirac cone surface and `M' TMs (red cross) in setup $S_{1}$ at $0\%$ strain. II) Dirac cone surface and `N' TMs (red cross) in setup $S_{1}$ at $10\%$ strain. III) Dirac cone surface and `M' TMs (red cross) in setup $S_{2}$ at $0\%$ strain. IV) Dirac cone surface and `P' TMs (red cross) in setup $S_{2}$ at $10\%$ strain. (b) The change in transmission due to deformation of constant energy surface (1 meV) due to strain in $S_{1}$ and $S_{2}$.} 

\label{norm_parameters}
\end{figure}

\subsection{Effect of strain on Dirac cone}
 
The linear regime in graphene corresponds to the energy close to the Dirac points. Therefore, we analyze the effect of strain on Dirac cone to understand the cause of piezoresistance in graphene.\\
\indent Graphene undergoes elastic deformation up to $20\%$ of strain. Simultaneously, it is highly resistive to band gap opening and undergoes band opening beyond $23\%$ of strain along the zigzag direction. In contrast, band gap does not open with applied strain along the armchair direction. The Dirac point shifts its position from K-points due to the applied strain~\cite{Pereira2009,Ni2009}. In this work, we restrict our discussion to $10\%$ of uniaxial strain.\\ 
\indent The shift in Dirac points $DP_{1}$ and $DP_{2}$ with respect to $K_{1}$ and $K_{2}$ at $10\%$ strain along armchair and zigzag directions are illustrated in Fig.~\ref{ac_0_dp1_all} and Fig.~\ref{zz_0_dp1_all} respectively. Table~\ref{table:DP} lists the relative shift between $K_{1}$ and $DP_{1}$, and $K_{2}$ and $DP_{2}$ with strain in reciprocal space. For strain along the armchair direction, Dirac points $DP_{1}$ and $DP_{4}$ move inside the Brillouin zone and Dirac points $DP_{2}$, $DP_{3}$, $DP_{5}$, and $DP_{6}$ move away from the Brillouin zone (Fig.~\ref{ac_0_dp1_all}). For strain along the zigzag direction, $DP_{1}$ and $DP_{4}$ move away from the Brillouin zone whereas the Dirac points $DP_{2}$ and $DP_{3}$ as well as $DP_{5}$ and $DP_{6}$ move closer to each other along the edge of Brillouin zone (refer to Fig.~\ref{zz_0_dp1_all}). We observe an identical response of $DP_{1}$ and $DP_{4}$ in $1^{st}$ Brillouin zone due to symmetry (Fig.~\ref{dp_sub}). Similarly, $DP_{2}$, $DP_{3}$, $DP_{5}$ and $DP_{6}$ are symmetrical and show identical response to applied strain. Shifting of these two sets of Dirac points with respect to the K-points are exactly equal and opposite. Therefore, we conclude that analysis of Dirac cones at $DP_{1}$ and $DP_{2}$ are sufficient to understand the strain response of all other Dirac cones.\\
\indent In addition to shifting of the Dirac points, strain induces distortion in Dirac cones. Application of uniaxial strain changes the Dirac cone into an elliptical cone as illustrated in Fig.~\ref{dc_strain}. All six Dirac cones have identical deformation for an applied strain. The dimension of the Dirac cones depend only on the magnitude of applied strain and poisson ratio. Thus, the tight binding parameters do not affect the shape of the Dirac cones.\\
\indent In unstrained graphene, each of the six Dirac cones contribute $1/3$ to the $1^{st}$ Brillouin zone. Effectively, two Dirac cones are present inside the Brillouin zone. The same is true in the case of uniaxially strained graphene shown in Fig.~\ref{dp_sub}. In Fig.~\ref{ac_0_dp1_all}, Dirac cones at $DP_{1}$ and $DP_{4}$ are present inside the $1^{st}$ Brillouin zone. Similarly, in Fig.~\ref{ac_0_dp1_all}, only one-half of the Dirac cones at $DP_{2}$, $DP_{3}$, $DP_{5}$ and $DP_{6}$ are present inside the $1^{st}$ Brillouin zone. Effectively, only two Dirac cones lie inside the $1^{st}$ Brillouin zone for strain along armchair direction or zigzag direction.

    \begin{table}
    \caption{Relative shift of Dirac points with respect to K-points with strain (refer Fig.\ref{dp_sub}).}
	\centering
	\begin{tabular}{|c|c|c|c|c|} 

		\hline
		$\mathbf{Strain}$  & $\mathbf{S_{1}}$ & $\mathbf{S_{1}}$ & $\mathbf{S_{2}}$ & $\mathbf{S_{2}}$ \\		
		
		$\mathbf{i\%}$ & $\mathbf{\alpha_{1}^{i}(\AA^{-1})}$ & $\mathbf{\alpha_{2}^{i}(\AA^{-1})}$ & $\mathbf{\beta_{1}^{i}(\AA^{-1})}$ &  $\mathbf{\beta_{2}^{i}(\AA^{-1})}$ \\		
		\hline\hline

		1\% & 0.0022 & -0.0022 & -0.0069 & 0.0069 \\
		\hline	
		5\% & 0.0137 & -0.0137 & -0.0401 & 0.0401 \\
		\hline
		10\% & 0.0337 & -0.0337 & -0.0995 & 0.0994\\
		\hline	
	\end{tabular}

\label{table:DP}
\end{table}

 \subsection{Physics of gauge factor variation}
 
Piezoresistance in a ballistic conductor is due to the change in transmission with applied strain. The change in transmission is primarily due to the change in band structure. Owing to the zero band gap, piezoresistance is only due to deformation of the Dirac cones with applied strain (refer to Fig.~\ref{dc_strain}).\\
\indent For armchair strained graphene, the modes in Dirac cones at $DP_{1}$ and $DP_{4}$ for electron transport along a particular direction (let's say +k) is equal to the modes along -k direction due to symmetry. Therefore, the total number of modes along +k direction in the $1^{st}$ Brillouin zone is equal to the sum of modes along `+k' and `-k' directions in Dirac cones at $DP_{1}$ or $DP_{2}$. Similarly, modes for strain along zigzag direction is equal to the sum of modes along +k direction and -k direction in any one of the Dirac cones at $DP_{2}$, $DP_{3}$, $DP_{5}$ and $DP_{6}$ (refer to Fig.~\ref{zz_0_dp1_all}).\\
\indent In Fig.~\ref{dc_strain}, the deformation of Dirac cones are identical for same magnitude of strain along armchair and zigzag directions. Furthermore, the separation between TMs are same in setup $S_{1}$ and $S_{2}$ of zigzag and armchair directions at same magnitude of strain. Consequently, we obtain the same LGF and TGF along zigzag and armchair directions. \\
\indent  Figure~\ref{modes} illustrate TMs at 1 meV constant energy surface of a Dirac cone at $0\%$ and $10\%$ strain in $S_{1}$ and $S_{2}$ setups. The parallel dotted lines represent the TMs whereas the red cross marks represent the particular energy modes.
The number of modes in $S_{1}$ reduces from `M' to `N' as the strain is increased from $0\%$ to $10\%$ along zigzag directions (refer to Fig.~\ref{modes}.I and Fig.~\ref{modes}.II). The net reduction in mode density is due to decrease in length of the minor axis of the constant energy surface and increase in separation between the TMs due to decrease in width. Thus, we see a gradual decrease in transmission with strain in $S_{1}$ as shown in Fig.~\ref{norm_transmission}. Whereas in setup $S_{2}$, due to an increase in major axis length of constant energy surface and reduction in separation of TMs, a significant increase in transmission is seen as shown in Fig.~\ref{norm_transmission}. The pattern of change in transmission at any other energy in linear regime remains identical as the major and minor axis length of the constant energy surface are proportional at a definite value of strain. Thus, the total change in transmission in setup $S_{1}$ and $S_{2}$ follows the same trend as the one shown in Fig.~\ref{norm_transmission} for E=1meV. The ratio of change in transmission at 1 meV energy in $S_{1}$ and $S_{2}$ is $\approx$ 10 times. This explains the sizeable variation in LGF and TGF value obtained in our simulations.

  \begin{table}
	\centering
	     \caption{Axes length ($\AA^{-1}$) of Dirac cone with varied strain at E=1 meV}
	\begin{tabular}{|c|c|c|c|c|} 
		\hline
		$\mathbf{Strain}$  & $\mathbf{DP_{1}(S_{1})}$ & $\mathbf{DP_{1}(S_{1})}$ & $\mathbf{DP_{2}(S_{2})}$ & $\mathbf{DP_{2}(S_{2})}$ \\		
		$\mathbf{i\%}$ & $\mathbf{(L_{x})}$ & $\mathbf{(L_{y})}$ & $\mathbf{(L_{x})}$ &  $\mathbf{(L_{y})}$ \\		
		\hline
		0\% & 0.000361 & 0.000362 & 0.000361 & 0.000362 \\
		\hline	
		5\% & 0.000413 & 0.000354 & 0.000354 & 0.000413 \\
		\hline
		10\% & 0.000495 & 0.000347 & 0.000347 & 0.000495\\
		\hline	
	\end{tabular}
\label{table:DC}
\end{table}

\section{Conclusion} \label{section4}

In this paper, we investigated the longitudinal and transverse piezoresistance in suspended graphene in ballistic regime. Utilizing parametrized tight binding Hamiltonian from ab initio calculations along with Landauer quantum transport formalism, we devised a methodology to evaluate the piezoresistance effect in 2D materials. We computed the longitudinal and transverse gauge factors of graphene along armchair and zigzag directions in the linear elastic limit ($0\%$-$10\%$). The longitudinal and transverse gauge factor values were identical along armchair and zigzag directions. Our model predicted a significant variation ($\approx 1000\% $ increase) in the magnitude of transverse gauge factor compared to longitudinal gauge factor along with sign inversion. The calculated value of longitudinal gauge factor is $\approx 0.3$ whereas the transverse gauge factor is $\approx -3.3$. We rationalized our prediction using deformation of Dirac cone and change in separation between transverse modes due to an applied uniaxial strain, leading to a change in resistance. Our results show a linear relationship between resistance and applied strain in longitudinal and transverse configurations. Thus, implementation of suspended graphene as a strain sensor in ballistic regime seems feasible. Owing to the low thickness, graphene strain pressure gauges have a very high sensitivity per unit area. Thus, the piezoresistance in transverse configuration can be extremely useful for pressure sensing. Based on our results, we suggested a suspended graphene based nano pressure sensor. The results obtained herein may serve as a template for piezoresistance effect of graphene in ballistic regime in nano electromechanical systems.

\begin{acknowledgments}

We thank Prof S.D. Mahanti for insightful discussion. We acknowledge Mr. Pankaj Priyadarshi, Mr. Debasis Das and Mr. Tejas R. Naik for their valuable contribution in preparing this manuscript. This work is supported by CEN Phase-2 and  NNETRA projects (Spons/EE/SG-3/2016) at IIT Bombay.  The Research and Development work undertaken in the project
under the Visvesvaraya Ph.D. Scheme of Ministry of Electronics and Information Technology, Government of India, being
implemented by Digital India Corporation (formerly Media
Lab Asia). This work was also supported by the Science and
Engineering Research Board (SERB) of the Government of
India under Grant No. EMR/2017/002853.

\end{acknowledgments}

\appendix

\section{{Strained tight binding parameters} \label{app1}}
The tight binding parameters used to obtain the dispersion relation of graphene at varied strain along zigzag and armchair directions are as follows:
\begin{table}[h]
	\caption{Hopping parameters for strain along zigzag direction}
	\centering
	\begin{tabular}{|c|c|c|} 
		\hline
		$\mathbf{Strain(\%)}$  & $\mathbf{t_{1}=t_{3}~(in~eV)}$ & $\mathbf{t_{2}~(in~eV)}$ \\	
		
		\hline\hline
		0\% & 2.60 & 2.60  \\
		\hline
		5\% & 2.36 & 2.68 \\
		\hline
		10\% & 2.08 & 2.75 \\
		\hline	
	\end{tabular}
	
	\label{table:zz}
\end{table}

\begin{table}[h]
	\caption{Hopping parameters for strain along armchair directions}
	\centering
	\begin{tabular}{|c|c|c|} 
		
		\hline
		$\mathbf{Strain(\%)}$  & $\mathbf{t_{1}=t_{3}~(in~eV)}$ & $\mathbf{t_{2}~(in~eV)}$ \\	
		
		\hline\hline
		0\% & 2.60 & 2.60  \\
		\hline
		5\% & 2.55 & 2.24 \\
		\hline
		10\% & 2.50 & 1.88 \\
		\hline	
	\end{tabular}
	
	\label{table:ac}
\end{table}
\section{{Expression for transmission of graphene} \label{app2}}

In Appendix \ref{app2}, we systematically derive the expression for transmission of graphene sheet starting with expression of mode density of a graphene's subband using the band counting method.\\
\indent The mode density of a parabolic subband with minima at energy $E_{1}$, is given by:
         \begin{equation}
         \begin{split}
          M(E) = \Theta(E-E_{1})
         \end{split}
        \label{A1}
         \end{equation}
The function `$\Theta(E)$' denotes a unit step function. Similarly, mode density of three different parabolic subbands, each having energy minima at $E_{1}$, $E_{2}$ and $E_{3}$ is given by: 
        \begin{equation}
        \begin{split}
         M(E) = \sum_{p=1}^{3}\Theta(E-E_{p})
        \end{split}
        \label{A2}
        \end{equation}
        
        \begin{figure}[h]
        {\includegraphics[scale=0.1]{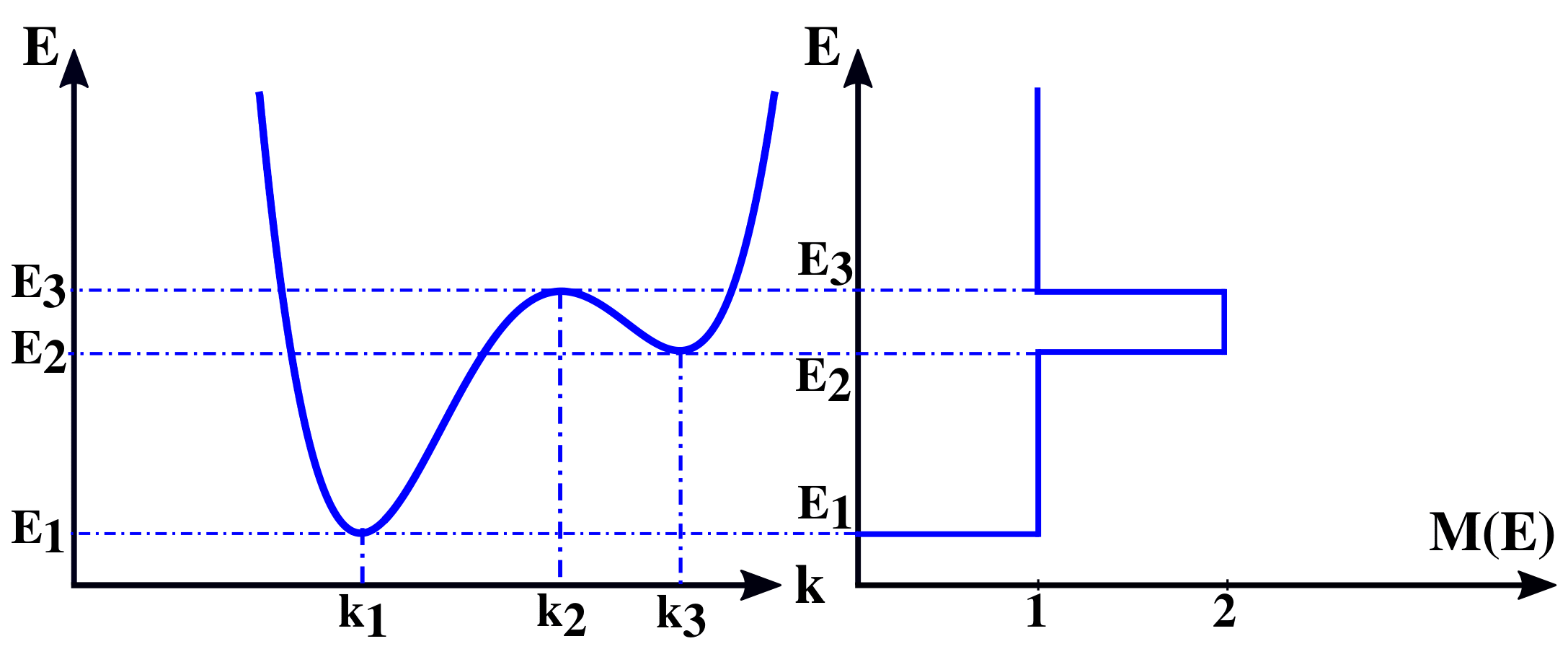}} 
        \caption{A subband with minima at $E_{1}$, $E_{2}$ and maxima at $E_{3}$, and its corresponding mode density function.}  
        \label{sb}
        \end{figure}
                  
Likewise, the mode density of a subband with energy minima at $E_{1}$ and $E_{2}$, and maxima at $E_{3}$(Fig.~\ref{sb}) is given by:
      \begin{equation}
        \begin{split}
          M(E) = \sum_{p=1}^{2}\Theta(E-E_{p})-\Theta(E-E_{3})
            \end{split}
          \label{A3}
         \end{equation}
Therefore, mode density of a subband with `p' minimas and `q' maximas is given by:
    \begin{equation}
       \begin{split}
    M(E)=\sum_{p=1}^{p}\Theta(E-E_{p})-\sum_{q=1}^{q}\Theta(E-E_{q})
      \end{split}
    \label{A4}
    \end{equation}
In general, Eq.~\eqref{A4} represents the mode density of a conduction band. The bandstructure of graphene is symmetric about $k_{x}$ and $k_{y}$ axes as well as the energy axis. Thus, a subband along armchair or zigzag direction has the same number of modes along forward and backward directions. The conduction and valence bands are symmetric about the energy axis. Thus, the collective equation for mode density of a graphene subband (say `$k_{\perp}$') consisting of the conduction band and valence band is given by:
\begin{equation}
   \begin{split}
    M^{i}_{k_\perp}(E) = \sum_{p=1}^{p} \Theta(E\mp E^{i}_{p}) - \sum_{q=1}^{q} \Theta(E\mp E^{i}_{q})
    \end{split}
    \label{A5}
\end{equation}
where, `$i$' is the strain percentage along transport direction. The negative signs in Eq.~\eqref{A5} are used for conduction band whereas the positive signs are used for valence band. We use Eq.~\eqref{A5} to systematically calculate the mode density function of the graphene sheet.\\
\indent The mode density of all TMs inside the segment $\Delta k$ (refer to Fig~\ref{sub-band}) is given by:
 \begin{equation}
  M^{i'}_{k_{\perp}}(E) = {\frac{M^{i}_{k_{\perp}}(E)*\Delta k}{\delta k}}
  \label{A6}
  \end{equation}
     
  $\Delta k=\Delta k_{y}$ for transport along the armchair direction and $\Delta k=\Delta k_{x}$ for transport along the zigzag direction. $\delta k$ is the separation between two adjacent TMs and has value of $\frac{2\pi}{L^{i}_{cs}}$ along both directions.
  
 Therefore, Eq.\eqref{A6} simplifies into: 
 
     \begin{equation}
      \begin{split}
       M^{i'}_{k_{\perp}}(E) & = \frac{M^{i}_{k_{\perp}}(E)*\Delta k*L^{i}_{cs}}{2\pi}
      \end{split}
     \label{A7}
      \end{equation}
      
Therefore, transmission per unit cross-sectional length by TMs present inside the segment $\Delta k$ containing subband ($k_{\perp}$) is given by: 
      \begin{equation}
      T^{i}_{k_{\perp}}(E)   = \frac{M^{i}_{k_{\perp}}(E)*\Delta k}{2\pi}
      \label{A8}
      \end{equation}
    From Eq.~\eqref{A8}, the prefactor($P^{i}_{k_{\perp}}$) is given by:
    \begin{equation}
    \begin{split}
    P^{i}_{k_{\perp}} & = \frac{\Delta k}{2\pi} \\ 
    \end{split}
    \label{A9}
    \end{equation}
 Equivalently, $T^{i}_{k_{\perp}}(E)$ can be written as:
    \begin{equation}
    \therefore T^{i}_{k_{\perp}}(E) =   P^{i}_{k_{\perp}}*M^{i}_{k_{\perp}}(E)\\
    \label{A10}
    \end{equation}
Finally, the total transmission per unit cross-sectional length of graphene sheet at $i\%$ strain when $j$ segments are used is given by:
    \begin{equation}
    \begin{split}
    T^{i}(E)=\sum_{k_{\perp}=1}^{j}T^{i}_{k_{\perp}}(E)\\
     \end{split}
     \label{A11}
     \end{equation} 
\section{{Expression for gauge factor}\label{app3}}
The expression for gauge factor is given by:
\begin{equation}
 GF =\frac{R^{i}-R^{0}}{R^{0}\varepsilon} \\
\label{B1}
\end{equation}
where $R^{i}$ is resistance of graphene sheet at $i\%$ strain and $R^{0}$ is resistance of graphene sheet at $0\%$ strain.
The resistance $R^{i}$ at $i\%$ strain is given by:       

\begin{equation}
R^{i} = \frac{r^{i}}{L^{i}_{cs}}
  = \frac{r^{i}}{L_{cs}*(1+\nu \varepsilon)}
\label{B2}
\end{equation} 
where $r^i$ is the gradient of voltage and current density in the linear regime at strain $i\%$. We put the value of Eq.~\eqref{B2} in Eq.~\eqref{B1} and obtain the final expression of GF:
\begin{equation}
\begin{split}
 GF = \frac{1}{\varepsilon}\Bigg[{\frac {r^{i}}{r^{0} (1+\nu \varepsilon)} -1\Bigg]}
 \end{split}
 \label{B3}
\end{equation}

\bibliography{reference01}

\end{document}